\documentclass[
reprint,
superscriptaddress,
amsmath,
amssymb,
aps,
prl,
]{revtex4-1}

\usepackage{bm}
\usepackage{seqsplit}
\usepackage{xcolor}
\usepackage{graphicx}
\usepackage{amsmath}
\usepackage{url}

\newcommand{\dm}[1]{\textcolor{black}{#1}}
\newcommand{\dmrev}[1]{\textcolor{black}{#1}}

\begin{document}

\title{Yeast condensin acts as a transient intermolecular crosslinker in entangled DNA}

\author{Filippo Conforto}
\thanks{joint first author}
\affiliation{School of Physics and Astronomy, University of Edinburgh, Peter Guthrie Tait Road, Edinburgh, EH9 3FD, UK}

\author{Antonio Valdes}
\thanks{joint first author}
\affiliation{Chair of Biochemistry and Cell Biology, Theodor-Boveri-Institute, Julius Maximilian University of Würzburg, 97074 Würzburg, Germany}

\author{Willem Vanderlinden}
\affiliation{School of Physics and Astronomy, University of Edinburgh, Peter Guthrie Tait Road, Edinburgh, EH9 3FD, UK}

\author{Davide Michieletto}
\email{davide.michieletto@ed.ac.uk}
\affiliation{School of Physics and Astronomy, University of Edinburgh, Peter Guthrie Tait Road, Edinburgh, EH9 3FD, UK}
\affiliation{MRC Human Genetics Unit, Institute of Genetics and Cancer, University of Edinburgh, Edinburgh EH4 2XU, UK}
\affiliation{International Institute for Sustainability with Knotted Chiral Meta Matter (WPI-SKCM$^2$), Hiroshima University, Higashi-Hiroshima, Hiroshima 739-8526, Japan}

\begin{abstract}
Structural-Maintenance-of-Chromosome (SMC) complexes such as condensins organise the folding of chromosomes. However, their role in modulating the entanglement of DNA and chromatin is not fully understood. To address this question, we perform single molecule and bulk characterisation of yeast condensin in entangled DNA. First, we discover that yeast condensin can proficiently bind double-stranded DNA through its hinge domain, in addition to its heads. Through bulk microrheology assays we then discover that physiological concentrations of yeast condensin increase both the viscosity and elasticity of dense solutions of $\lambda$-DNA suggesting that condensin acts as a crosslinker in entangled DNA, stabilising entanglements rather than resolving them and contrasting the popular theoretical picture where SMCs purely drive the formation of segregated, bottle-brush-like chromosome structures. We further discover that the presence of ATP fluidifies the solution -- likely by activating loop extrusion -- but does not recover the viscosity measured in the absence of protein. Finally, we show that the observed rheology can be understood by modelling SMCs as transient crosslinkers in bottle-brush-like entangled polymers. Our findings help us to understand how SMCs affect the dynamics and entanglement of genomes. 
\end{abstract}

\maketitle

\section*{Introduction}

Among the most important processes orchestrating chromosome folding in both interphase and mitosis is the formation of loops, performed by structural-maintenance-of-chromsome (SMC) complexes, such as cohesin, condensin and SMC5/6.
Although these complexes perform loop extrusion \textit{in vitro}, the extent to which loop extrusion affects genome organisation and dynamics \textit{in vivo} is poorly understood. 

Alternative models to loop extrusion are able to explain experimental observations, both \textit{in vivo} and \textit{in vitro}. For instance, bridging-induced phase separation (BIPS) model can explain the formation of clusters, or condensates, of yeast cohesin in the presence of DNA. The loop capture model can explain the topological trapping of a DNA plasmid by a condensin that is loaded on a tethered linear DNA. Both these models rely on the fact that SMCs can ``bridge'' inter-chromosomal DNA, i.e. can simultaneously bind two dsDNA segments that do not belong to the same DNA molecule. While there is evidence of SMC bridging for cohesin, such evidence is less abundant for condensin. In fact, intermolecular bridging is not at all envisioned in loop extrusion models, as SMCs are envisioned to ``reel in'' DNA \textit{in cis}. 
Mixed models, whereby SMCs perform an \textit{effective} loop extrusion by bridging DNA segments have also been proposed and can capture some puzzling evidence, for instance the formation of Z-loops and the bypassing of large obstacles bound to DNA, or the observation that condensin can make steps larger than its own size. 

SMCs are expected to have a significant impact on the dynamics of chromosomes in cells, however it is challenging to precisely quantify it experimentally. The prediction from most computational and theoretical works is that loop extrusion performed by SMCs will compact, segregate, fluidify and even unknot chromatin, implying that SMCs should speed up chromosome dynamics. However, indirect evidence obtained by single-particle tracking of H2B and chromosome loci suggest that rapid cohesin depletion yields a speed up of chromosome dynamics and nucleosome motion. Moreover, live-cell studies have consistently demonstrated that cohesin constrains chromatin dynamics. In fission yeast, disrupting loop factors increases locus mobility, a finding mirrored in mammalian mESCs upon acute cohesin depletion. High-resolution tracking in human cells further revealed this constraint operates at the nucleosome level, reducing the internal fluidity of euchromatic domains, implying the exact opposite of current theoretical and computational models, i.e. that cohesin slows down chromosome dynamics. 

Thus, there is a clear disconnect among (i) \textit{in vitro} single molecule evidence displaying SMC loop extrusion,  (ii) theoretical work suggesting SMC loop extrusion should drive chromosome compaction and speed up genome dynamics and (iii) \textit{in vivo} evidence suggesting that SMCs slow down chromosome dynamics. 

In this work, we aim to bridge the gap between existing evidence and rectify this disconnect. To do this, we perform bulk and single molecule assays on yeast condensin on \textit{entangled} DNA \textit{in vitro}. This is different from any previous work \emph{in vitro} as they mostly focused on tethered DNA or dilute conditions. Instead, to understand the role of SMCs \textit{in vivo}, we argue that we must study their behaviour in physiologically dense DNA solutions. 

The key discovery of this work is that we find evidence supporting the claim that most existing computational and theoretical models are incomplete. Indeed, we observe that yeast condensin is a proficient intermolecular bridge and acts as ``thickening'' agent in entangled solutions of $\lambda$-DNA. Importantly, we also discover that this ``thickening'' is mostly loop extrusion independent. We conclude our paper by suggesting an alternative model for SMC as ``sticky loop extruders'' which can perform both loop extrusion but also intermolecular bridging, thus forming transient cross-linking in dense DNA solutions. Our results contribute to understanding the action of SMC proteins in physiologically crowded and entangled environments such as those of the cell nucleus.

\section*{Materials and Methods}

\subsection*{Protein expression and purification}
Wild-type (WT) and Q-loop condensin holocomplexes were expressed from two 2$\mu$-based high copy plasmids transformed into \textit{S. cerevisiae}. 
Purification of holocomplexes has been performed as in Ref.~\cite{Shaltiel2022} (see SI for more details). 
Expression of yeast Smc2 residues 396-792 and yeast Smc4 residues 555-951 was induced from pET-MCN vectors in bacteria. Smc2 (396-792) with an N-terminal (His)6-TEV-tag and untagged Smc4 (555-951) we co-expressed and purified by Ni-Sepharose 6FF (GE Healthcare), Resource Q (GE Healthcare) and Superdex 200 GL 10/300 column (GE Healthcare) (see SI for full details).

\subsection*{Electrophoretic mobility shift assay (EMSA)}
The 6-FAM labeled 50-bp dsDNA was prepared by annealing two complementary DNA oligos (Merck, \seqsplit{5'-6-FAM-GGATACGTAACAACGCTTATGCATCGCCGCCGCTACATCCCTGAGCTGAC-3'}; \seqsplit{5'-GTCAGCTCAGGGATGTAGCGGCGGCGATGCATAAGCGTTGTTACGTATCC-3'}) in annealing buffer (50 mM Tris-HCl pH 7.5, 50 mM NaCl) at a concentration of 50 $\mu$M in a temperature gradient of 0.1 C/s from 95$^\circ$C to 4$^\circ$C. The EMSA reaction was prepared with a constant DNA concentration of 10 nM and the indicated concentrations of purified protein in binding buffer (50 mM Tris-HCl pH 7.5, 50 mM KCl, 125 mM NaCl, 5mM MgCl2, 5\% Glycerol, 1 mM DTT). After 10 min incubation on ice, free DNA and DNA-protein complexes were resolved by electrophoresis for 1.5 hr at 4 V/cm, on 0.75\% (w/v) TAE-agarose gels at 4$^\circ$C. 6-FAM labeled dsDNA was detected directly on a Typhoon FLA 9,500 scanner (GE Healthcare) with excitation at 473 nm with LPB (510LP) filter setting. 

\subsection*{Fluorescence Polarisation}
Fluorescence polarization (FP) experiment was performed by mixing 20 nM of the 6-FAM labeled 50 bp dsDNA (see Methods EMSA) with series of protein concentrations, ranging from 0.03125 $\mu$M to 32 $\mu$M, in FP buffer (25 mM Tris-HCl pH 7.5, 100 mM NaCl, 5 mM MgCl2, 1 mM DTT, 0.05\% Tween20, 0.05 mg/ml BSA). The mix was incubated for 30 min at room temperature in order to attain equilibrium. Immediately thereafter, fluorescence polarization was recorded using 485 nm and 520 nm excitation and emission filter on a Tecan SPARK Microplate reader. The change in fluorescence polarization was then plotted as mean values of three independent replicates and the dissociation constant determined.

\subsection*{AFM imaging}
Atomic Force Microscopy (AFM) was performed on poly-L-lysine-coated mica. Linear dsDNA of 500 bp was generated by PCR from pUC19 plasmid using primers 5'-AGAGCAACTCGGTCGCCGCATA (forward) and 5'-GCTTACCATCTGGCCCCAGTGC (reverse)). We mixed 0.5 ng/$\mu$L DNA and 10 nM WT condensin in aqueous buffer (50 mM Tris-HCl pH = 7.5, 25 mM NaCl, 5 mM MgCl2, 1mM DTT, 1mM ATP) and incubated at room temperature for 15 seconds before deposition. Deposition of the sample onto poly-L-lysine coated mica was done by dropcasting. After surface adsorption for 15 s, the sample was rinsed using milliQ water (20 mL) and subsequently dried using a gentle stream of filtered N2 gas. For imaging the sample we used a Nanowizard 4 XP AFM (JPK, Berlin, Germany) in tapping mode; image processing was done using MountainSPIP software (see SI).

\subsection*{Microrheology}
For microrheology experiments, we mixed 5 $\mu$l of 500 ng/$\mu$l $\lambda$DNA with 1 $\mu$l of 2 $\mu$M yeast condensin (WT or Q), $1$ $\mu$l of 10x condensin reaction buffer (Tris-HCl pH 7.5 500 mM, NaCl 250 mM, MgCl2 50 mM, DTT 10 mM), 1 $\mu$l of 10 mM ATP and 1 $\mu$l of 2 $\mu$m PEGylated polystyrene beads (Polyscience). We loaded the sample into a 100 $\mu$m thick sample chamber comprising a microscope slide, 100 $\mu$m layer of double-sided tape and a cover slip.  We recorded movies on a Nikon Eclipse Ts2 microscope with a 20x objective and Orca Flash 4.0 CMOS camera (Hamamatsu) for 2 minutes at $\sim 100$ fps on a 1024x1024 field of view, resulting in about 500 tracks per condition. Particle tracking was done using trackpy and in-house code was used to process the tracks into MSD and complex modulus following Ref.~\cite{Mason2000}.

\subsection*{Molecular Dynamics Simulations}
Entangled DNA solutions were modelled as semiflexible Kremer-Grest linear polymers with $N=500$ beads of size $\sigma = 10$ nm. The beads interact with each other via a truncated and shifted Lennard-Jones potential and adjacent beads are connected by FENE springs. The persistence length of the polymers is $l_p = 5 \sigma = 50$ nm and the volume fraction of the solution is around 5 \%.
After thorough equilibration (see SI), the polymers are loaded with $N_{SMC} = \{5, 25\}$ SMCs and then either let to loop extrude as in Ref.~\cite{Conforto2024}, or otherwise left on the loaded state to mimic conditions with no ATP. Each SMC is decorated with patches that have an attractive interaction with the DNA beads, modelled by a Morse potential with a maximum depth of $25 k_BT$, which is comparable with the heads (and hinge) binding affinity $\Delta G \approx - k_B T \log{(k_D)}$ where $k_D \simeq 0.1$ $\mu$M. The simulation is performed in LAMMPS with custom-made fixes that update the position of active loop extruders (\url{https://git.ecdf.ed.ac.uk/taplab/smc-lammps}). Specifically, our loop extrusion algorithm performs a geometry check before updating the position of the SMCs in order to preserve the topology of the system. We then track the dynamics of the polymers when the SMCs are only loaded (no loops) and when allowed to make large loops via loop extrusion. In the latter case, the polymers start to resemble bottle-brushes. At the same time, we perform Green-Kubo calculations of the stress relaxation function, i.e. we compute the autocorrelation of the off-diagonal components of the stress tensor, in order to obtain a measure of viscoelasticity in the system under different conditions (see SI for more details).

\section*{Results}

\begin{figure*}
        \centering
        \includegraphics[width=0.9\textwidth]{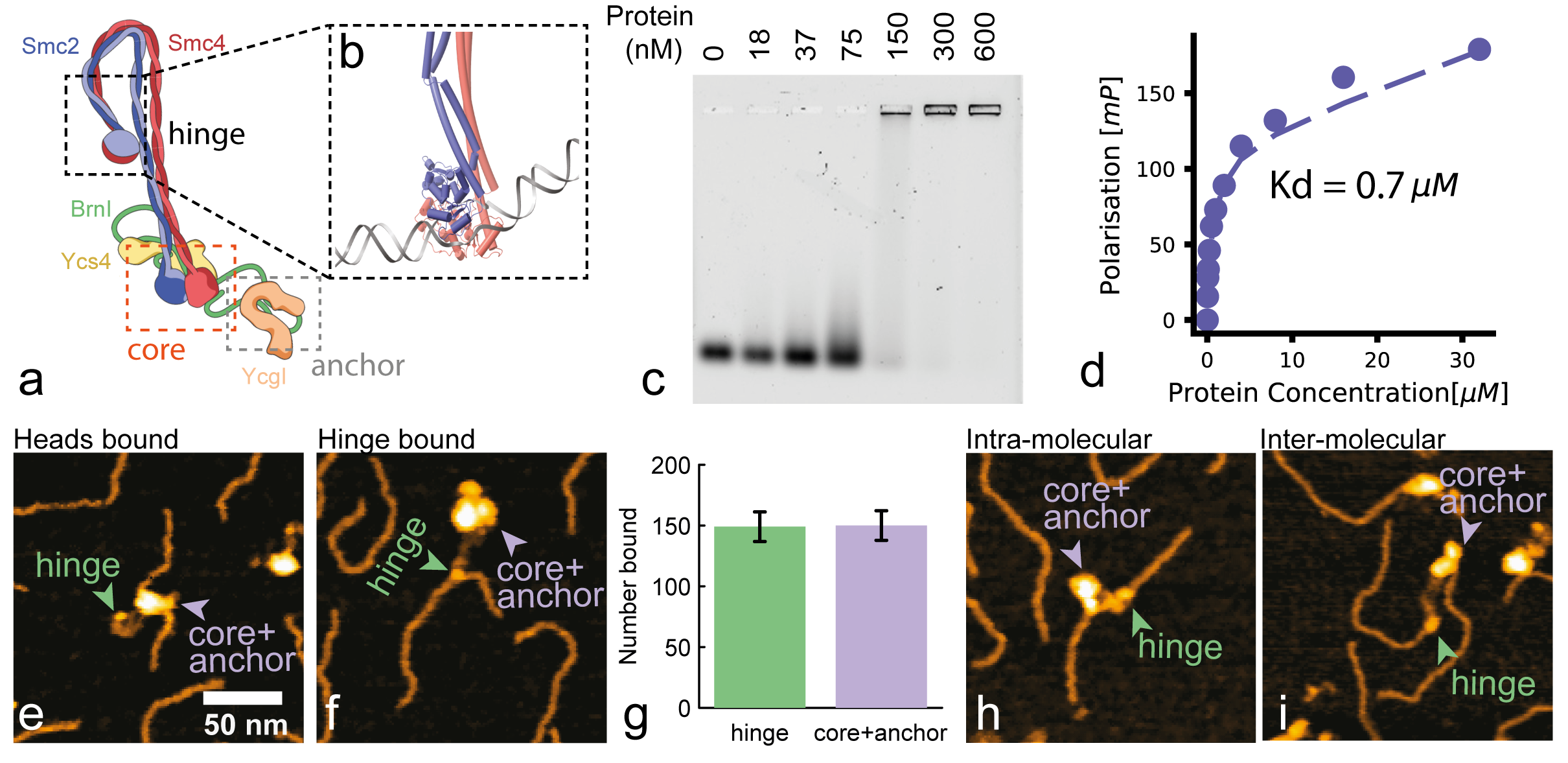}
        \caption{ 
        \textbf{Yeast condensin proficiently binds dsDNA through its hinge domain.} \textbf{a.} Cartoon structure of yeast condensin; different domains are highlighted. \textbf{b.} AlphaFold3 model of structural interaction between the hinge domain and a segment of dsDNA with a small ssDNA bubble in the middle. \textbf{c.} EMSA showing significant binding of the hinge domain (SMC2:K841-L698, SMC4:Q646-F865) to a 25 bp dsDNA oligo \textit{in vitro} with an estimated $k_D \simeq 0.075-0.15$ $\mu$M. \textbf{d.} Fluorescence polarisation assay done with the hinge domain mixed with fluorescently-labelled 50 bp dsDNA oligo and yielding $k_D = 0.7$ $\mu$M.     
        \textbf{e-f.} Representative AFM topographs of (e) head-bound and (f) hinge-bound condensin-DNA complexes. Green and lilac arrowheads indicate hinges and core+anchor domains, respectively. \textbf{g.} Quantification of relative hinge and heads bound complexes.  Error bars reflect counting statistics $\sqrt{N_i}/N_{total}$. $N_{total} = 299$. \textbf{h-i.} Representative AFM topographs of (h) intra-molecular and (i) inter-molecular condensin-DNA complexes.}
        \label{fig:fig2}
\end{figure*}

\subsection*{Yeast condensin can form intermolecular bridges by binding dsDNA at its hinge domain}

First, to better understand the role of condensin in dense solutions of DNA, we decided to investigate different binding modes of yeast condensin to dsDNA. Condensin binds dsDNA through both its ``anchor'' domain (BrnI-YcgI) and its ``core'' subcomplex (SMC heads + Ycs4), however there is no direct evidence of dsDNA binding by any other condensin domain (fig.~\ref{fig:fig2}a). Thus, we sought computational evidence for an additional binding site by scanning through AlphaFold3 structures. We found a model that predicted an interaction between the SMC2/SMC4 hinge and a dsDNA oligomer containing a small ssDNA bubble (fig.~\ref{fig:fig2}b). Motivated by this prediction, we performed electrophoretic mobility shift assay (EMSA) and observed a clear shift when the hinge domain (SMC2:K841-L698, SMC4:Q646-F865) was mixed with a 50 bp dsDNA segment (fig.~\ref{fig:fig2}c), with an estimated binding affinity of $K_d \simeq 0.075-0.15$ $\mu$M.  

This measurement was further supported by fluorescence polarization (FP), where yeast condensin hinge was mixed with a fluorescently-labelled dsDNA oligo, albeit we measured a larger binding constant $K_d \simeq 0.7$ $\mu$M  (fig.~\ref{fig:fig2}d). Interestingly, these $K_d$ values are comparable to -- if not smaller than -- the binding constants of the YcgI-BrnI (anchor) complex to DNA, i.e. $K_d \simeq 1.7$ $\mu$M and of the ``core'' subcomplex (SMC heads + Ycs4) $K_d \simeq 0.1 -0.2$ $\mu$M, both measured from \textit{Chaetomium thermophilum}. Arguably, both EMSA and FP potentially underestimate the true $K_d$ because they employ short dsDNA oligos, which are not the natural substrate for these protein complexes; however, they convincingly demonstrate that the hinge is a proficient dsDNA binding site potentially as good as the core/anchor subcomplex.

Motivated by these measurements, we decided to visualise dsDNA binding by the whole yeast condensin holocomplex in single molecule experiments. We mixed yeast condensin holocomplex with a 500 bp dsDNA segment, deposited it on mica and observed it using Atomic Force Microscopy (AFM, see Methods). We observed that yeast condensin displays different modes of binding: through its core+anchor complex, hinge, or both (fig.~\ref{fig:fig2}e-f,h-i). When the holocomplex bound DNA through its core+anchor domains, we also observed a severe kink in the dsDNA molecule, in agreement with the cryo-EM structure and the ``safety-belt'' model (see fig.~\ref{fig:fig2}e). On the other hand, we observed no deformation of the substrate DNA when the hinge was bound to it (see fig.~\ref{fig:fig2}f). 

Surprisingly, out of 299 molecules analysed, 149 were bound by the hinge and 150 bound by the core+anchor domains; when both were bound, we considered that both heads and hinge were bound. This result confirms that hinge and core+anchor domains display similar binding affinities to dsDNA. While this is broadly in line with the bulk EMSA and FP assays, it is an aspect of SMC biophysics that has been overlooked and it is not accounted for in any of the existing models (they all start from core+anchor domains bound to DNA and an unbound hinge). Interestingly, we also observed a significant number of intra and inter-molecular bridging, whereby two segments of DNA belonging to \textit{different} molecules are simultaneously bound by the core+anchor and hinge. This evidence suggests that yeast condensin may be a proficient ``bridging'' protein, as observed \textit{in vitro} for yeast cohesin. 

Finally, we argue that while the thermodynamics of condensin domains binding to DNA may be similar, the kinetics of binding/unbinding may be very different for the two domains, e.g. due to their local flexibility. We hypothesise that the ``safety-belt'' anchoring mechanism at the YcgI-BrnI domain may be very stable (small $k_{off}$ and small $k_{on}$), whilst the kinetics at the hinge may be faster (large $k_{off}$ and large $k_{on}$). In turn, the ratio of the on/off-rates give similar equilibrium constant $K_d$. This reasoning could explain both the strong structural evidence for the ``safety-belt'' anchoring and the elusive DNA-hinge interaction as well as the potential for forming transient inter-molecular bridges, loops and cross-links.

\subsection*{Condensin acts as transient crosslinker in entangled DNA even during loop extrusion}

\begin{figure*}[t!]
    \centering
    \includegraphics[width=0.9\textwidth]{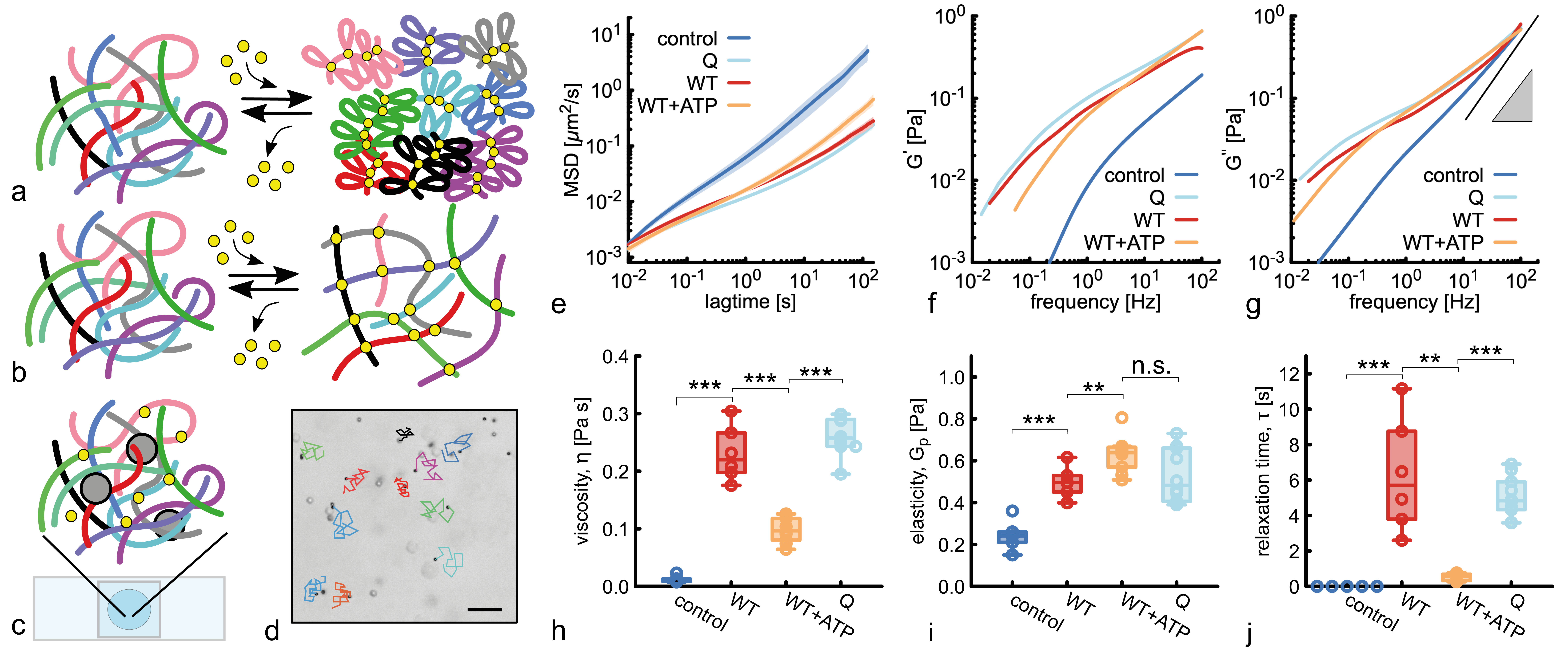}
    \caption{ \textbf{Microrheology reveals that SMCs can form intermolecular bridges in entangled DNA.}\textbf{a-b} Sketches of our two hypotheses: \textbf{a.} if condensin performed loop extrusion (intramolecular contacts only), we would expect a solution of entangled linear DNA to be converted into one made of bottle-brush-like polymers, reducing both entanglement and viscoelasticity. \textbf{b.} If condensin performed DNA-bridging (intermolecular contacts), we would expect transient crosslinks. \textbf{c.} The sample made of $\lambda$DNA, condensin, reaction buffer and passive tracers is mixed, incubated and then pipetted in a closed chamber. \textbf{d.} Snapshot of the field of view showing the tracers and short example trajectories (scale bar 20 $\mu$m). \textbf{e.} Mean squared displacement (MSD) of the tracer beads for wild type yeast condensin (WT) in presence and absence of ATP and for a catalytically dead (Q) mutant. For all samples in this figure, DNA concentration is $250$ ng/$\mu$L (or 7.8 nM of $\lambda$DNA) and protein concentration is $0.2$ $\mu$M, i.e. about 25 SMCs per DNA molecule. \textbf{f-g.} Elastic ($G^\prime$, \textbf{f}) and viscous ($G^{\prime \prime}$, \textbf{g}) complex moduli obtained from the MSDs through the generalised Stokes Einstein relation. \textbf{h.} Zero-shear viscosity, obtained from the long time behaviour of the MSD. P-values in the plot: $* < 0.05$, $** < 0.01$, $*** < 0.001$. The p-value between WT and Q mutant is 0.13, and hence non significant. \textbf{i.} Elasticity $G^\prime_p$ obtained from the elastic modulus measured at 100 Hz. \textbf{j.} Relaxation time $\tau_R$, obtained as the inverse of the smallest frequency at which $G^\prime$ and $G^{\prime \prime}$ intersect.}
    \label{fig:fig1}
\end{figure*}

To understand the effect of SMC intermolecular bridging observed in the previous section, we decided to assess the effect of SMCs on the rheology of entangled DNA. We hypothesised that if condensin was mainly performing intra-molecular loop extrusion, we would observe a significant decrease in entanglements and a consequent fluidification of the solution, i.e. a \textit{decrease} in viscosity. This hypothesis is in line with current models of SMCs on chromosomes and DNA, where loop extrusion is envisaged to drive the formation of bottle-brush-like structures. On the other hand, intermolecular bridging would give rise to gel-like networks, whereby DNA-DNA entanglements would be stabilised by condensin bridges. These entangled networks of DNA are expected to display larger viscosity and elasticity than equally dense DNA solutions without SMC (fig.~\ref{fig:fig1}a,b). 

To quantitatively measure condensin effect on the viscoelasticity of DNA solutions, we prepared samples of entangled $\lambda$-DNA (48'502 bp) at around 12 times the overlap concentration ($c = 250$ ng/$\mu$l $= 7.8$ nM, $c^* = 20$ ng/$\mu$l), and mixed it with 0.2 $\mu$M of either wild type (WT) yeast condensin or a catalytically dead (Q-loop) mutant that cannot perform loop extrusion. These conditions represent a dense, entangled solution of long monodisperse DNA whereby each polymer has, on average, 10-20 SMCs loaded onto it (or 1 SMC every 5 kbp). We also included 2 $\mu$m-sized PEG-passivated polystyrene tracer beads (we obtained similar results with different beads' sizes, see SI) and adjusted buffer conditions to those used to observe loop extrusion in single molecule assays. After incubation at 37$^\circ$C for 5 minutes, we added 1mM ATP and loaded $5$ $\mu$l of sample onto a chamber made of a glass slide and coverslip, kept apart by a 100 $\mu$m spacer, and visualised it under an inverted microscope (fig.~\ref{fig:fig1}c). We then performed microrheology, i.e. recorded videos of the passive tracers moving in the solution and extracted their mean squared displacement (MSD) $\delta^2 r(t) = \langle [ \bm{r}(t+t_0) - \bm{r}(t_0) ]^2\rangle$, where the average is performed over beads, initial times $t_0$, sample location and at least three independent replicates (see fig.~\ref{fig:fig1}e).

According to most current models, SMCs should compact DNA by performing loop extrusion and thus decrease the viscosity of the entangled solution. In our experiment, this fluidification would manifest itself as an increase in effective diffusion coefficient of the tracer beads and an absence of subdiffusive behaviour. On the contrary, we observed the opposite: a significant decrease in the mobility of the beads and an increase in their subdiffusive regime for both WT and Q-loop condensin and in both presence and absence of ATP (fig.~\ref{fig:fig1}e). To quantify the elastic and viscous response of the fluid at different timescales we transformed the MSDs into elastic ($G^\prime(\omega)$) and viscous ($G^{\prime \prime}(\omega)$) complex moduli via the generalised Stokes-Einstein relation. In Figs.~\ref{fig:fig1}f-g, one can appreciate that the presence of WT and Q-loop condensin significantly affects the shape of $G^\prime(\omega)$ and $G^{\prime \prime}(\omega)$. More specifically, the control displays a purely viscous behaviour with little sign of inflection in $G^{\prime \prime}(\omega)$; on the contrary, the samples with SMCs display at least one intersection between the two complex moduli (see SI). This entails that the fluid's response is elastic-dominated at short timescales (large $\omega$) and liquid-dominated at long timescales (small $\omega$). Finally, we also observe that SMCs induce a significant increase in both elasticity and viscosity of the samples across all frequencies.

We compute the zero-shear viscosity of the samples $\eta = k_B T/(3 \pi a D)$, where $a = 2$ $\mu$m is the size of the beads and $D$ the large-time diffusion coefficient obtained from the MSDs. We note that adding SMCs induce a 20-30-fold increase in viscosity in all samples, with the increase being more pronounced for WT and Q-loop mutant (fig.~\ref{fig:fig1}h). Interestingly, the sample with WT SMC and ATP displays the smallest change ($\sim 10$-fold), \dm{which may point to a fluidification effect of loop extrusion}. 

We can also compute the large-$\omega$ elasticity $G^{\prime}_p$ and relaxation time $\tau_R$ of these viscoelastic fluids by evaluating $G^{\prime}_p=G^\prime(\omega=\textrm{100 Hz})$ and $G^\prime(1/\tau_R) = G^{\prime \prime}(1/\tau_R)$, respectively. The former ($G^{\prime}_p$, fig.~\ref{fig:fig1}i) suggest that the short-time elastic behaviour is significantly stiffer for SMC samples, regardless of whether there is ATP or not. Despite this observation, all samples display $G^{\prime}_p < 1$ Pa, implying that they are very soft. On the other hand, the latter ($\tau_R$, fig.~\ref{fig:fig1}j) suggest that samples with WT condensin and ATP behave like liquids on shorter timescales (smaller $\tau_R$) than the ones without ATP or with the Q-loop mutant which remain solid-dominated for longer times, up to tens of seconds.  

Our observations suggest that SMCs may form transient cross-links between DNA molecules that are proximal in 3D space. \dm{Further, our results suggest that WT in the absence of ATP behaves similarly to the Q-loop mutant. This suggests that both proteins bind DNA similarly and that the Q-loop mutation does not affect the crosslinking ability of SMC. However, in the presence of ATP, we observe a significant ``fluidification'', which we argue is an effect of loop extrusion. However, we find that this fluidification is not strong enough to fully counteract the transient SMC-mediated crosslinking.} 

\subsection*{Coarse grained MD simulations of ``sticky'' SMCs \dm{captures} the behaviour seen in bulk and single-molecule assays}

\begin{figure*}[t!]
    \centering
    \includegraphics[width=0.9\textwidth]{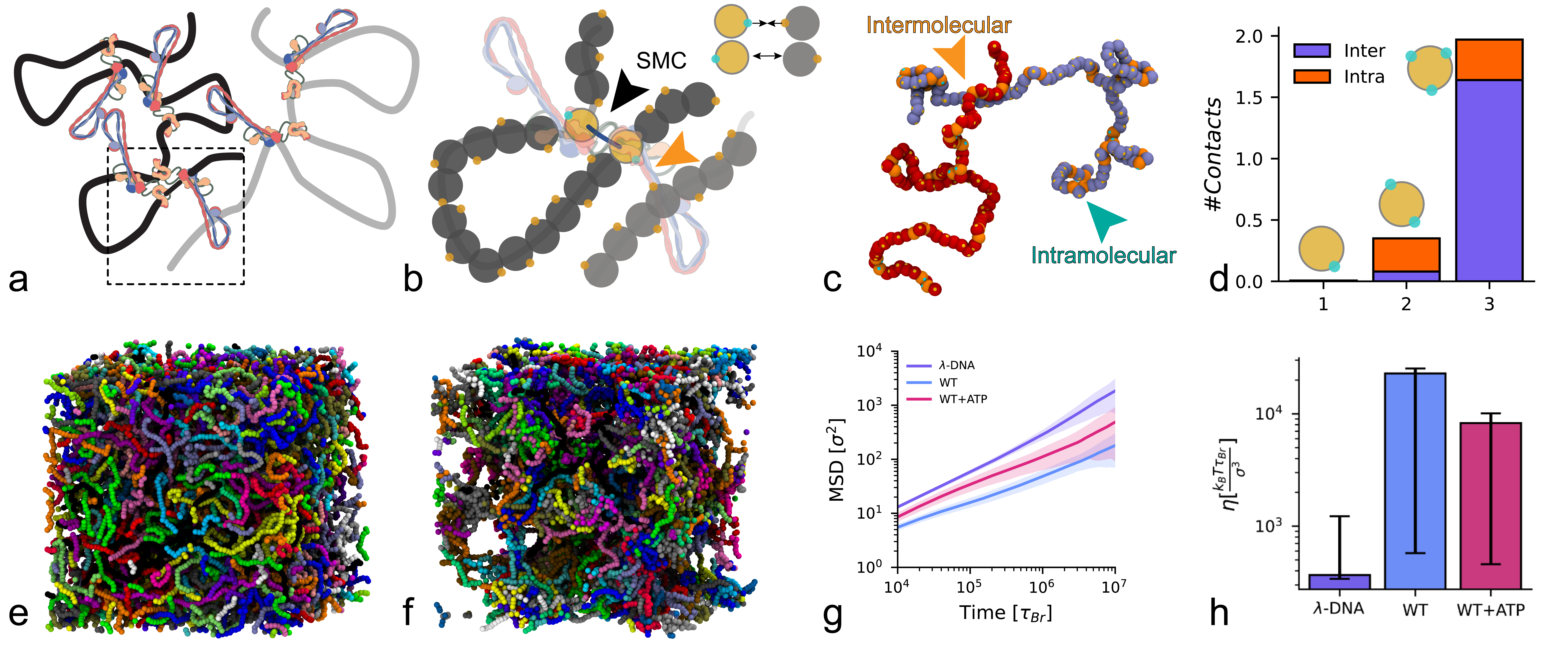}
    \caption{
    \textbf{MD simulations of sticky SMCs model the thickening.} \textbf{a.} Sketch of DNA with loops formed by SMCs. \textbf{b.} Bead-spring polymer modelling of the region in the dashed box showing the correspondence between patches (cyan) and hinge domain. \textbf{c.} Snapshot from simulations, highlighting intramolecular and intermolecular interactions stabilised by the patches. \textbf{d} Average number of contacts as a function of the number of patches on the beads. \textbf{e-f.} Snapshots of the simulation box in two cases:  (\textbf{e}) in equilibrium with no SMC and (\textbf{f}) after loading 50 sticky SMCs per polymer and allowing them to extrude loops. \textbf{g.} Average Mean Squared Displacement (MSD) of the polymers' center of mass (standard deviation shaded) for the control case ($\lambda$-DNA) compared with the cases with SMC but no extrusion (WT) and the case with SMC allowed to extrude loops (WT+ATP). \textbf{h.} Viscosity computed from the stress-relaxation function (see SI) for the three cases in \textbf{g.} Notice that with ATP, the system is more fluid, in line with experiments.
    }
    \label{fig:fig3}
\end{figure*}

Motivated by the observations in the previous Sections we decided to test a simple coarse-grained model of ``sticky'' loop extruders. Specifically, we performed Molecular Dynamics (MD) simulations of entangled linear DNA under the action of SMCs that display small patches that can bind to DNA polymers (see fig.~\ref{fig:fig3}a-b). Briefly, we modelled DNA molecules as Kremer-Grest bead-spring polymers~\cite{Kremer1990} at fixed density and in the entangled regime, corresponding to 5\% volume fraction. After equilibrating the system, we randomly loaded, on average, 5 SMCs per polymer and allowed them to form both DNA loops \textit{in cis} and inter-molecular bridges through their patches (Fig.~\ref{fig:fig3}a-c). Our SMC model is different from most models in the literature as we allow the SMCs to do both, form intra/inter molecular bridges through their patches and also form loops through extrusion whilst preserving the polymer topology (see SI for full details).

To account for the formation of SMC clusters we also explored the effect of having multiple DNA binding sites on each SMC bead: on average, less than one contact per SMC complex was seen for patches per SMC bead, while 2 contacts (mostly inter-molecular) were seen for $N_p = 3$ patches per SMC bead (fig.~\ref{fig:fig3}d). This implies that only one third of all SMC patches were bound to DNA at any one time and corresponds to the case in which there are two clustered, or stacked, SMCs per loop. In this scenario, the two SMCs are bound to the same DNA by their core+anchor domains, and have two ``free'' hinges that can make inter or intra molecular contacts with other DNA segments (see fig.~\ref{fig:fig3}a-b).  

We observed that once the SMCs are loaded, the system qualitatively displayed remarkable clustering (see snapshots fig.~\ref{fig:fig3}e-f). By computing the mean squared displacement of the centre of mass of the polymers, i.e. $\delta^2 r (t) = \langle \left[ \bm{r}_{CM}(t + t_0) - \bm{r}_{CM}(t) \right]^2 \rangle$, we discovered that they also displayed a significantly slower dynamics (fig.~\ref{fig:fig3}g-h). 
More specifically, to compare our simulations with the experiments we performed two sets of simulations: (i) SMCs are bound to DNA and cannot loop extrude (case with no ATP or Q-loop mutant in experiments) and (ii) SMCs are allowed to extrude loops (case with ATP in experiments). The MSDs displayed in Fig.~\ref{fig:fig3}g show that the dynamics of the polymers in presence of extrusion is faster than the no-extrusion case, in agreement with experiments. We argue that this result is explained by the fact that in the latter case the polymers were forming bottle-brush-like organisations that reduced the overall entanglements and sped up their dynamics~\cite{Conforto2024,Chan2024pnas}. We further computed the stress relaxation function $G(t)$ through the autocorrelation of the out-of-diagonal components of the stress tensor~\cite{Ramirez2010,Lee2009a} (see SI\dm{, fig. S6}) and computed the zero-shear viscosity of the simulated fluid. In line with the microrheology, the DNA solution with SMCs that cannot perform loop extrusion display a 20-fold increase in viscosity. Allowing the SMC to loop extrusion only reduces the viscosity by a small ($\sim$2-fold) factor but the dominant effect remains the transient ``gelling'' of the DNA entanglements. 



\section*{Discussion and conclusions} 
In summary, in this paper we have provided experimental and computational evidence that SMCs, and specifically yeast condensin, can stabilise inter-molecular interactions in solutions of dense DNA.

First, we used biochemical assays and AFM to uncover that the hinge domain of yeast condensin is a proficient dsDNA binding site (fig.~\ref{fig:fig2}a-d). Unexpectedly, we observed that it binds as strongly as condensin heads, which are well-known DNA binding sites from structural studies~\cite{Shaltiel2022,Kschonsak2017}. \dm{We also note that although AlphaFold3 predicted the hinge domain binding to a ssDNA bubble, our experimental data clearly shows that it can bind both ss and dsDNA (see EMSA and FP in fig.~\ref{fig:fig2}c-d, fig.~\ref{fig:fig2}e-f, and SI, \dm{fig S1 and S2}). Interestingly, we also note that previous experiments suggested that condensin stepping can undertwist DNA, which is consistent with the formation of a small ssDNA bubble (5-6 bp for $\Delta Tw \simeq -0.5$)~\cite{Martinez-Garcia2022}. It is therefore tempting to hypothesise that SMC binding and stepping may itself induce the formation of a single-stranded bubble on dsDNA. Finally, by performing AFM imaging we report visual evidence that SMCs can simultaneously bind dsDNA through heads and hinge domains forming both intra and inter-molecular contacts (fig.~\ref{fig:fig2}h-i). }

\dm{In light of this evidence, we reasoned that if yeast condensin was to be introduced in a dense and entangled solution of DNA, it would mediate inter-molecular bridges. However, this hypothesis would be at odds with a large fraction of the current loop extrusion models which posit that SMCs mostly perform intra-chain loop extrusion and no bridging, leading to mostly unentangled, bottle-brush-like chromosome structures~\cite{Alipour2012,Goloborodko2016,Sanborn2015a,Kim2023}. We therefore decided to test the action of yeast condensin on entangled DNA using microrheology. Specifically, we performed experiments with $\lambda$-DNA at volume fraction comparable to that of DNA in yeast cells, i.e. $12 \textrm{Mbp} \times [\pi (0.34 \textrm{nm/bp}) (2.5 \textrm{nm})^2]/(4 \mu m^3) \simeq 2\%$ or about 3 mg/ml~\cite{Fosado2023Fluidification}. We decided to use 
0.25 mg/ml of $\lambda$-DNA at low ionic conditions, 25 mM NaCl, which yields an effectively larger DNA diameter of about 12 nm~\cite{Rybenkov1993DNAdiameter}; in turn, these conditions yield an entangled solution of $\lambda$-DNA at an effective $\sim 1\%$ volume fraction.} 

We reasoned that if pure intra-chain loop extrusion was the dominant mode of action of SMCs, we would observe a speed up of the dynamics of entangled DNA solutions, leading to a so-called ``thinning'' of the solution's viscosity (fig.~\ref{fig:fig1}a-b). In contrast, we consistently observed that adding yeast condensin to our solution of entangled DNA lead to so-called ``thickening'', i.e. an increase of the solution's viscosity and mirroring a slowdown of DNA dynamics (fig.~\ref{fig:fig1}e-g). This effect cannot be attributed to the mere presence of additional protein in the solution because (i) we observe increase in elasticity, implying the formation of DNA crosslinks and (ii) we observed thinning with different proteins (e.g. with IHF in Ref.~\cite{Fosado2023Fluidification}). \dm{Our microrheology data also reveals that adding ATP partially recovers the DNA dynamics. \dmrev{Since we excluded that this effect is due to ATP itself (see SI, fig.~S3)} we thus concluded that the partial fluidification observed in the presence of ATP and WT protein is likely due to ATP-driven loop extrusion partially counter-acting intermolecular crosslinking (fig.~\ref{fig:fig1}h). Finally, we discover that the SMC Q-loop mutant, despite its inability to hydrolise ATP, can also form intermolecular crosslinks, as strong as the wild type condensin (fig.~\ref{fig:fig1}i-j).}

To connect the rheology and single-molecule observations, we concluded this work by performing MD simulations where we modelled SMCs as ``sticky'' proteins that can both, stabilise dynamic intermolecular cross-links and form loops. This computational model yielded results in line with what observed experimentally (fig.~\ref{fig:fig3}a-b,g-h).

\begin{figure}[t!]
    \centering
    \includegraphics[width=0.4\textwidth]{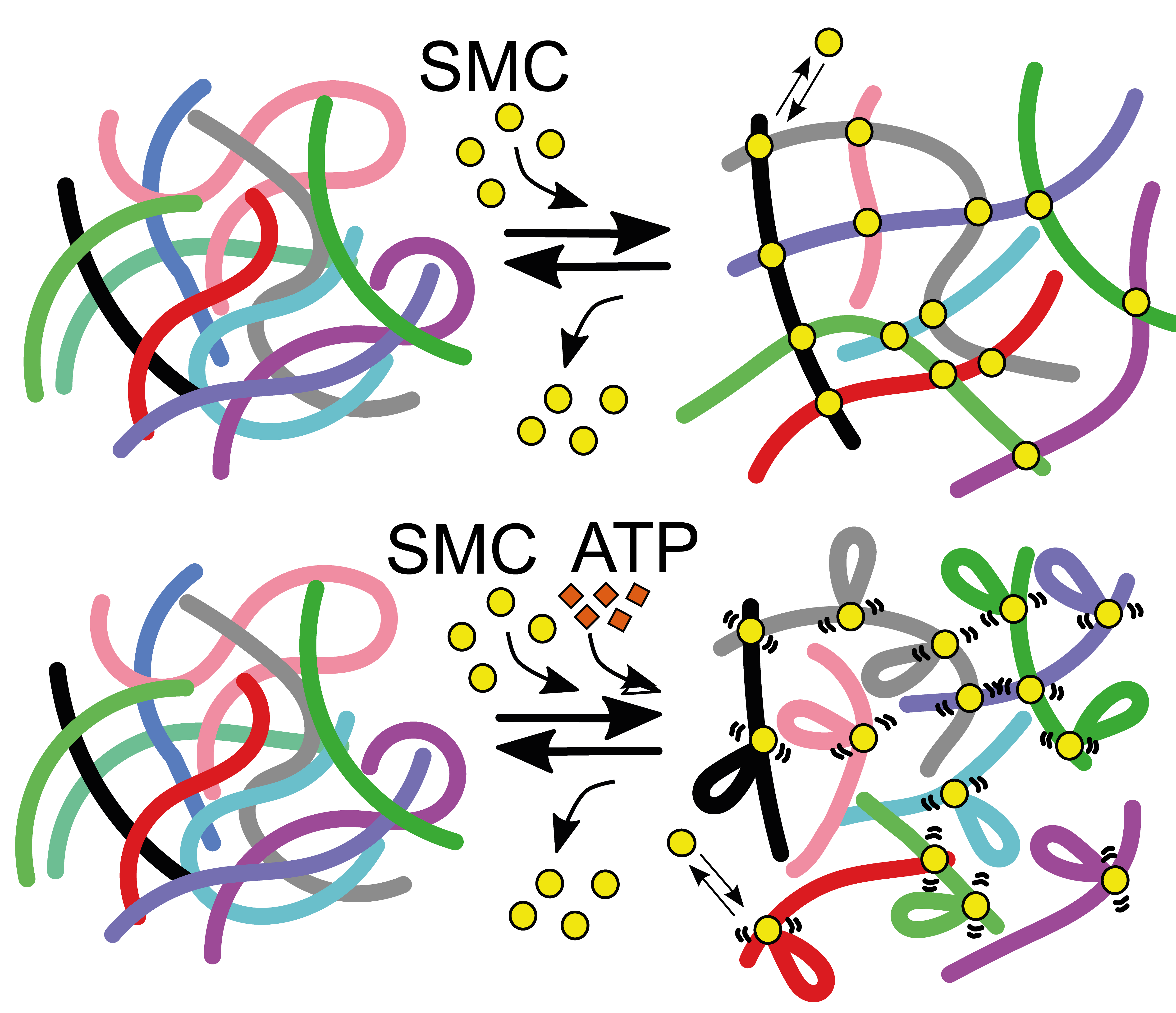}
    \caption{\textbf{SMCs act as transient intermolecular crosslinkers that can perform intramolecular loop extrusion in presence of ATP.} (top) SMCs loaded on DNA form transient intermolecular bridges by simultaneously binding dsDNA molecules through their heads and hinge domains. (bottom) ATP-driven loop extrusion competes with intermolecular bridging and speeds up DNA dynamics.}
    \label{fig:finalsketch}
\end{figure}

In light of this, we therefore argue that in our \textit{in vitro} experiments, SMCs do not exclusively form intra-chain loops (as predicted by loop extrusion models), but also form inter-molecular transient crosslinks, in turn affecting the solution's entanglements and viscoelasticity (fig.~\ref{fig:finalsketch}). 
Our model is in line with the ``bridging-induced'' phase separation behaviour observed in yeast cohesin~\cite{Ryu2021,Brackley2013pnas}, \dm{and the evidence that condensin can slow down chromatin dynamics in mitosis and interphase~\cite{Hibino2024,Nozaki2023}; it is also in agreement with the role of condensin in sequestering repetitive DNA in the nucleolus~\cite{Hult2017}}. By having identified the hinge as an additional dsDNA binding domain, our model of SMC acting as both intermolecular crosslinker and intramolecular loop extruder can naturally explain other models, e.g. the  ``loop capture''~\cite{Tang2023,Uhlmann2025} and ``inter-molecular loop-extrusion''~\cite{BONATO20215544,Bonato2025} models. 

\dm{The overall picture is that SMCs form a dynamic and reversible mesh of weakly cross-linked polymers, whereby the crosslinks themselves may be mobile if SMCs are performing loop extrusion (fig.~\ref{fig:finalsketch}). In fact, we argue that this system is physically similar to so-called slide-ring gels, where polymers in solution thread through ring-like molecules that can form crosslinks and slide along the chains~\cite{sapsford2025topologically}}.

\dm{Finally, we should highlight that our microrheology experiments are the first \textit{in vitro} evidence that condensin forms intermolecular, transient crosslinks in entangled DNA solutions. Rheology measurements on DNA solutions offer a clear quantification of the impact of SMCs on entangled DNA. Since our experiments are performed at around $\sim$ 1\% volume fraction, we believe that they are closer to physiological concentrations than current single molecule assays, e.e. DNA tethering or optical and magnetic tweezers~\cite{Ganji2018,Murayama2018}.} The effects uncovered in this work are therefore expected to be physiologically relevant and could in fact explain the puzzling observations from single molecule tracking \textit{in vivo}, whereby depletion of cohesin typically induces a speed up of chromatin dynamics in interphase~\cite{Nozaki2017,Gabriele2022,Nozaki2023} and depletion of condensin speeds up nucleosome dynamics during metaphase~\cite{hibino2024single}. At the same time, our work can explain the role of condensin in stiffening chromosomes through bridging~\cite{Sun2018}.    

To conclude, we argue that SMCs's role in regulating genome organisation and dynamics may be more multifaceted and complex than previously thought. Our experiments suggest that SMC intermolecular bridging is a dominant mechanism of action on entangled DNA and that intramolecular loop extrusion may potentially play a minor contribution. \dm{Additionally, the relative weight of these contributions may be modulated across the cell cycle by partner proteins. In the future, we argue that microrheology will be an ideal assay to test the presence of partner proteins and additional co-factors in a physiologically relevant condition of DNA density.} 

\section{Acknowledgements}
DM acknowledges the Royal Society and the European Research Council (grant agreement No 947918, TAP) for funding. The authors also acknowledge the contribution of the COST Action Eutopia, CA17139.  For the purpose of open access, the author has applied a Creative Commons Attribution (CC BY) licence to any Author Accepted Manuscript version arising from this submission. The authors thank Markus Hassler and Christian H\"{a}ring for comments and feedback on the manuscript.

\section{Author contributions}
FC: Conceptualisation (contributing); Software (lead); Formal analysis (equal); Investigation (equal); Visualization (contributing); Writing (contributing). AV: Conceptualization (lead); Methodology (equal); Investigation (equal). WV: Conceptualization (lead); Methodology (equal); Investigation (equal); Formal analysis (equal); Writing (equal). DM: Funding acquisition (lead); Supervision (lead); Conceptualization (lead); Methodology (equal); Investigation (equal); Formal analysis (equal); Visualisation (lead); Writing -- original draft (lead); Writing -- review and editing (lead).


\begin{thebibliography}{15}%
\makeatletter
\providecommand \@ifxundefined [1]{%
 \@ifx{#1\undefined}
}%
\providecommand \@ifnum [1]{%
 \ifnum #1\expandafter \@firstoftwo
 \else \expandafter \@secondoftwo
 \fi
}%
\providecommand \@ifx [1]{%
 \ifx #1\expandafter \@firstoftwo
 \else \expandafter \@secondoftwo
 \fi
}%
\providecommand \natexlab [1]{#1}%
\providecommand \enquote  [1]{``#1''}%
\providecommand \bibnamefont  [1]{#1}%
\providecommand \bibfnamefont [1]{#1}%
\providecommand \citenamefont [1]{#1}%
\providecommand \href@noop [0]{\@secondoftwo}%
\providecommand \href [0]{\begingroup \@sanitize@url \@href}%
\providecommand \@href[1]{\@@startlink{#1}\@@href}%
\providecommand \@@href[1]{\endgroup#1\@@endlink}%
\providecommand \@sanitize@url [0]{\catcode `\\12\catcode `\$12\catcode
  `\&12\catcode `\#12\catcode `\^12\catcode `\_12\catcode `\%12\relax}%
\providecommand \@@startlink[1]{}%
\providecommand \@@endlink[0]{}%
\providecommand \url  [0]{\begingroup\@sanitize@url \@url }%
\providecommand \@url [1]{\endgroup\@href {#1}{\urlprefix }}%
\providecommand \urlprefix  [0]{URL }%
\providecommand \Eprint [0]{\href }%
\providecommand \doibase [0]{http://dx.doi.org/}%
\providecommand \selectlanguage [0]{\@gobble}%
\providecommand \bibinfo  [0]{\@secondoftwo}%
\providecommand \bibfield  [0]{\@secondoftwo}%
\providecommand \translation [1]{[#1]}%
\providecommand \BibitemOpen [0]{}%
\providecommand \bibitemStop [0]{}%
\providecommand \bibitemNoStop [0]{.\EOS\space}%
\providecommand \EOS [0]{\spacefactor3000\relax}%
\providecommand \BibitemShut  [1]{\csname bibitem#1\endcsname}%
\let\auto@bib@innerbib\@empty
\bibitem [{\citenamefont {Mason}(2000)}]{Mason2000}%
  \BibitemOpen
  \bibfield  {author} {\bibinfo {author} {\bibfnamefont {T.~G.}\ \bibnamefont
  {Mason}},\ }\href {\doibase 10.1007/s003970000094} {\bibfield  {journal}
  {\bibinfo  {journal} {Rheologica Acta}\ }\textbf {\bibinfo {volume} {39}},\
  \bibinfo {pages} {371} (\bibinfo {year} {2000})}\BibitemShut {NoStop}%
\bibitem [{\citenamefont {Harnett}\ \emph {et~al.}(2024)\citenamefont
  {Harnett}, \citenamefont {Weir},\ and\ \citenamefont
  {Michieletto}}]{Harnett2024}%
  \BibitemOpen
  \bibfield  {author} {\bibinfo {author} {\bibfnamefont {J.}~\bibnamefont
  {Harnett}}, \bibinfo {author} {\bibfnamefont {S.}~\bibnamefont {Weir}}, \
  and\ \bibinfo {author} {\bibfnamefont {D.}~\bibnamefont {Michieletto}},\
  }\href {\doibase 10.1039/D3SM00957B} {\bibfield  {journal} {\bibinfo
  {journal} {Soft Matter}\ }\textbf {\bibinfo {volume} {20}},\ \bibinfo {pages}
  {3980} (\bibinfo {year} {2024})}\BibitemShut {NoStop}%
\bibitem [{\citenamefont {Vanderlinden}\ \emph {et~al.}(2014)\citenamefont
  {Vanderlinden}, \citenamefont {Lipfert}, \citenamefont {Demeulemeester},
  \citenamefont {Debyser},\ and\ \citenamefont {{De
  Feyter}}}]{Vanderlinden2014}%
  \BibitemOpen
  \bibfield  {author} {\bibinfo {author} {\bibfnamefont {W.}~\bibnamefont
  {Vanderlinden}}, \bibinfo {author} {\bibfnamefont {J.}~\bibnamefont
  {Lipfert}}, \bibinfo {author} {\bibfnamefont {J.}~\bibnamefont
  {Demeulemeester}}, \bibinfo {author} {\bibfnamefont {Z.}~\bibnamefont
  {Debyser}}, \ and\ \bibinfo {author} {\bibfnamefont {S.}~\bibnamefont {{De
  Feyter}}},\ }\href {\doibase 10.1039/c4nr00022f} {\bibfield  {journal}
  {\bibinfo  {journal} {Nanoscale}\ }\textbf {\bibinfo {volume} {6}},\ \bibinfo
  {pages} {4611} (\bibinfo {year} {2014})}\BibitemShut {NoStop}%
\bibitem [{\citenamefont {Chen}\ \emph {et~al.}(2003)\citenamefont {Chen},
  \citenamefont {Weeks}, \citenamefont {Crocker}, \citenamefont {Islam},
  \citenamefont {Verma}, \citenamefont {Gruber}, \citenamefont {Levine},
  \citenamefont {Lubensky},\ and\ \citenamefont {Yodh}}]{Chen2003}%
  \BibitemOpen
  \bibfield  {author} {\bibinfo {author} {\bibfnamefont {D.~T.}\ \bibnamefont
  {Chen}}, \bibinfo {author} {\bibfnamefont {E.~R.}\ \bibnamefont {Weeks}},
  \bibinfo {author} {\bibfnamefont {J.~C.}\ \bibnamefont {Crocker}}, \bibinfo
  {author} {\bibfnamefont {M.~F.}\ \bibnamefont {Islam}}, \bibinfo {author}
  {\bibfnamefont {R.}~\bibnamefont {Verma}}, \bibinfo {author} {\bibfnamefont
  {J.}~\bibnamefont {Gruber}}, \bibinfo {author} {\bibfnamefont {A.~J.}\
  \bibnamefont {Levine}}, \bibinfo {author} {\bibfnamefont {T.~C.}\
  \bibnamefont {Lubensky}}, \ and\ \bibinfo {author} {\bibfnamefont {A.~G.}\
  \bibnamefont {Yodh}},\ }\href {\doibase 10.1103/PhysRevLett.90.108301}
  {\bibfield  {journal} {\bibinfo  {journal} {Physical Review Letters}\
  }\textbf {\bibinfo {volume} {90}},\ \bibinfo {pages} {108301} (\bibinfo
  {year} {2003})}\BibitemShut {NoStop}%
\bibitem [{\citenamefont {Fosado}\ \emph {et~al.}(2023)\citenamefont {Fosado},
  \citenamefont {Howard}, \citenamefont {Weir}, \citenamefont {Noy},
  \citenamefont {Leake},\ and\ \citenamefont
  {Michieletto}}]{Fosado2023Fluidification}%
  \BibitemOpen
  \bibfield  {author} {\bibinfo {author} {\bibfnamefont {Y.~A.~G.}\
  \bibnamefont {Fosado}}, \bibinfo {author} {\bibfnamefont {J.}~\bibnamefont
  {Howard}}, \bibinfo {author} {\bibfnamefont {S.}~\bibnamefont {Weir}},
  \bibinfo {author} {\bibfnamefont {A.}~\bibnamefont {Noy}}, \bibinfo {author}
  {\bibfnamefont {M.~C.}\ \bibnamefont {Leake}}, \ and\ \bibinfo {author}
  {\bibfnamefont {D.}~\bibnamefont {Michieletto}},\ }\href {\doibase
  10.1103/PhysRevLett.130.058203} {\bibfield  {journal} {\bibinfo  {journal}
  {Physical Review Letters}\ }\textbf {\bibinfo {volume} {130}},\ \bibinfo
  {pages} {058203} (\bibinfo {year} {2023})}\BibitemShut {NoStop}%
\bibitem [{\citenamefont {Kremer}\ and\ \citenamefont
  {Grest}(1990)}]{Kremer1990}%
  \BibitemOpen
  \bibfield  {author} {\bibinfo {author} {\bibfnamefont {K.}~\bibnamefont
  {Kremer}}\ and\ \bibinfo {author} {\bibfnamefont {G.~S.}\ \bibnamefont
  {Grest}},\ }\href {\doibase 10.1063/1.458541} {\bibfield  {journal} {\bibinfo
   {journal} {The Journal of Chemical Physics}\ }\textbf {\bibinfo {volume}
  {92}},\ \bibinfo {pages} {5057} (\bibinfo {year} {1990})}\BibitemShut
  {NoStop}%
\bibitem [{\citenamefont {Tubiana}\ \emph {et~al.}(2024)\citenamefont
  {Tubiana}, \citenamefont {Alexander}, \citenamefont {Barbensi}, \citenamefont
  {Buck}, \citenamefont {Cartwright}, \citenamefont {Chwastyk}, \citenamefont
  {Cieplak}, \citenamefont {Coluzza}, \citenamefont {{\v{C}}opar},
  \citenamefont {Craik}, \citenamefont {{Di Stefano}}, \citenamefont
  {Everaers}, \citenamefont {Fa{\'{i}}sca}, \citenamefont {Ferrari},
  \citenamefont {Giacometti}, \citenamefont {Goundaroulis}, \citenamefont
  {Haglund}, \citenamefont {Hou}, \citenamefont {Ilieva}, \citenamefont
  {Jackson}, \citenamefont {Japaridze}, \citenamefont {Kaplan}, \citenamefont
  {Klotz}, \citenamefont {Li}, \citenamefont {Likos}, \citenamefont
  {Locatelli}, \citenamefont {L{\'{o}}pez-Le{\'{o}}n}, \citenamefont {Machon},
  \citenamefont {Micheletti}, \citenamefont {Michieletto}, \citenamefont
  {Niemi}, \citenamefont {Niemyska}, \citenamefont {Niewieczerzal},
  \citenamefont {Nitti}, \citenamefont {Orlandini}, \citenamefont {Pasquali},
  \citenamefont {Perlinska}, \citenamefont {Podgornik}, \citenamefont
  {Potestio}, \citenamefont {Pugno}, \citenamefont {Ravnik}, \citenamefont
  {Ricca}, \citenamefont {Rohwer}, \citenamefont {Rosa}, \citenamefont {Smrek},
  \citenamefont {Souslov}, \citenamefont {Stasiak}, \citenamefont {Steer},
  \citenamefont {Su{\l}kowska}, \citenamefont {Su{\l}kowski}, \citenamefont
  {Sumners}, \citenamefont {Svaneborg}, \citenamefont {Szymczak}, \citenamefont
  {Tarenzi}, \citenamefont {Travasso}, \citenamefont {Virnau}, \citenamefont
  {Vlassopoulos}, \citenamefont {Ziherl},\ and\ \citenamefont
  {{\v{Z}}umer}}]{Tubiana2024}%
  \BibitemOpen
  \bibfield  {author} {\bibinfo {author} {\bibfnamefont {L.}~\bibnamefont
  {Tubiana}}, \bibinfo {author} {\bibfnamefont {G.~P.}\ \bibnamefont
  {Alexander}}, \bibinfo {author} {\bibfnamefont {A.}~\bibnamefont {Barbensi}},
  \bibinfo {author} {\bibfnamefont {D.}~\bibnamefont {Buck}}, \bibinfo {author}
  {\bibfnamefont {J.~H.}\ \bibnamefont {Cartwright}}, \bibinfo {author}
  {\bibfnamefont {M.}~\bibnamefont {Chwastyk}}, \bibinfo {author}
  {\bibfnamefont {M.}~\bibnamefont {Cieplak}}, \bibinfo {author} {\bibfnamefont
  {I.}~\bibnamefont {Coluzza}}, \bibinfo {author} {\bibfnamefont
  {S.}~\bibnamefont {{\v{C}}opar}}, \bibinfo {author} {\bibfnamefont {D.~J.}\
  \bibnamefont {Craik}}, \bibinfo {author} {\bibfnamefont {M.}~\bibnamefont
  {{Di Stefano}}}, \bibinfo {author} {\bibfnamefont {R.}~\bibnamefont
  {Everaers}}, \bibinfo {author} {\bibfnamefont {P.~F.}\ \bibnamefont
  {Fa{\'{i}}sca}}, \bibinfo {author} {\bibfnamefont {F.}~\bibnamefont
  {Ferrari}}, \bibinfo {author} {\bibfnamefont {A.}~\bibnamefont {Giacometti}},
  \bibinfo {author} {\bibfnamefont {D.}~\bibnamefont {Goundaroulis}}, \bibinfo
  {author} {\bibfnamefont {E.}~\bibnamefont {Haglund}}, \bibinfo {author}
  {\bibfnamefont {Y.~M.}\ \bibnamefont {Hou}}, \bibinfo {author} {\bibfnamefont
  {N.}~\bibnamefont {Ilieva}}, \bibinfo {author} {\bibfnamefont {S.~E.}\
  \bibnamefont {Jackson}}, \bibinfo {author} {\bibfnamefont {A.}~\bibnamefont
  {Japaridze}}, \bibinfo {author} {\bibfnamefont {N.}~\bibnamefont {Kaplan}},
  \bibinfo {author} {\bibfnamefont {A.~R.}\ \bibnamefont {Klotz}}, \bibinfo
  {author} {\bibfnamefont {H.}~\bibnamefont {Li}}, \bibinfo {author}
  {\bibfnamefont {C.~N.}\ \bibnamefont {Likos}}, \bibinfo {author}
  {\bibfnamefont {E.}~\bibnamefont {Locatelli}}, \bibinfo {author}
  {\bibfnamefont {T.}~\bibnamefont {L{\'{o}}pez-Le{\'{o}}n}}, \bibinfo {author}
  {\bibfnamefont {T.}~\bibnamefont {Machon}}, \bibinfo {author} {\bibfnamefont
  {C.}~\bibnamefont {Micheletti}}, \bibinfo {author} {\bibfnamefont
  {D.}~\bibnamefont {Michieletto}}, \bibinfo {author} {\bibfnamefont
  {A.}~\bibnamefont {Niemi}}, \bibinfo {author} {\bibfnamefont
  {W.}~\bibnamefont {Niemyska}}, \bibinfo {author} {\bibfnamefont
  {S.}~\bibnamefont {Niewieczerzal}}, \bibinfo {author} {\bibfnamefont
  {F.}~\bibnamefont {Nitti}}, \bibinfo {author} {\bibfnamefont
  {E.}~\bibnamefont {Orlandini}}, \bibinfo {author} {\bibfnamefont
  {S.}~\bibnamefont {Pasquali}}, \bibinfo {author} {\bibfnamefont {A.~P.}\
  \bibnamefont {Perlinska}}, \bibinfo {author} {\bibfnamefont {R.}~\bibnamefont
  {Podgornik}}, \bibinfo {author} {\bibfnamefont {R.}~\bibnamefont {Potestio}},
  \bibinfo {author} {\bibfnamefont {N.~M.}\ \bibnamefont {Pugno}}, \bibinfo
  {author} {\bibfnamefont {M.}~\bibnamefont {Ravnik}}, \bibinfo {author}
  {\bibfnamefont {R.}~\bibnamefont {Ricca}}, \bibinfo {author} {\bibfnamefont
  {C.~M.}\ \bibnamefont {Rohwer}}, \bibinfo {author} {\bibfnamefont
  {A.}~\bibnamefont {Rosa}}, \bibinfo {author} {\bibfnamefont {J.}~\bibnamefont
  {Smrek}}, \bibinfo {author} {\bibfnamefont {A.}~\bibnamefont {Souslov}},
  \bibinfo {author} {\bibfnamefont {A.}~\bibnamefont {Stasiak}}, \bibinfo
  {author} {\bibfnamefont {D.}~\bibnamefont {Steer}}, \bibinfo {author}
  {\bibfnamefont {J.}~\bibnamefont {Su{\l}kowska}}, \bibinfo {author}
  {\bibfnamefont {P.}~\bibnamefont {Su{\l}kowski}}, \bibinfo {author}
  {\bibfnamefont {D.~W.~L.}\ \bibnamefont {Sumners}}, \bibinfo {author}
  {\bibfnamefont {C.}~\bibnamefont {Svaneborg}}, \bibinfo {author}
  {\bibfnamefont {P.}~\bibnamefont {Szymczak}}, \bibinfo {author}
  {\bibfnamefont {T.}~\bibnamefont {Tarenzi}}, \bibinfo {author} {\bibfnamefont
  {R.}~\bibnamefont {Travasso}}, \bibinfo {author} {\bibfnamefont
  {P.}~\bibnamefont {Virnau}}, \bibinfo {author} {\bibfnamefont
  {D.}~\bibnamefont {Vlassopoulos}}, \bibinfo {author} {\bibfnamefont
  {P.}~\bibnamefont {Ziherl}}, \ and\ \bibinfo {author} {\bibfnamefont
  {S.}~\bibnamefont {{\v{Z}}umer}},\ }\href {\doibase
  10.1016/j.physrep.2024.04.002} {\bibfield  {journal} {\bibinfo  {journal}
  {Physics Reports}\ }\textbf {\bibinfo {volume} {1075}},\ \bibinfo {pages} {1}
  (\bibinfo {year} {2024})}\BibitemShut {NoStop}%
\bibitem [{\citenamefont {Plimpton}(1995)}]{Plimpton1995}%
  \BibitemOpen
  \bibfield  {author} {\bibinfo {author} {\bibfnamefont {S.}~\bibnamefont
  {Plimpton}},\ }\href {\doibase 10.1006/jcph.1995.1039} {\bibfield  {journal}
  {\bibinfo  {journal} {J. Comp. Phys.}\ }\textbf {\bibinfo {volume} {117}},\
  \bibinfo {pages} {1} (\bibinfo {year} {1995})}\BibitemShut {NoStop}%
\bibitem [{\citenamefont {Fudenberg}\ \emph {et~al.}(2016)\citenamefont
  {Fudenberg}, \citenamefont {Imakaev}, \citenamefont {Lu}, \citenamefont
  {Goloborodko}, \citenamefont {Abdennur},\ and\ \citenamefont
  {Mirny}}]{fudenbergFormationChromosomalDomains2016}%
  \BibitemOpen
  \bibfield  {author} {\bibinfo {author} {\bibfnamefont {G.}~\bibnamefont
  {Fudenberg}}, \bibinfo {author} {\bibfnamefont {M.}~\bibnamefont {Imakaev}},
  \bibinfo {author} {\bibfnamefont {C.}~\bibnamefont {Lu}}, \bibinfo {author}
  {\bibfnamefont {A.}~\bibnamefont {Goloborodko}}, \bibinfo {author}
  {\bibfnamefont {N.}~\bibnamefont {Abdennur}}, \ and\ \bibinfo {author}
  {\bibfnamefont {L.~A.}\ \bibnamefont {Mirny}},\ }\href {\doibase
  10.1016/j.celrep.2016.04.085} {\bibfield  {journal} {\bibinfo  {journal}
  {Cell Reports}\ }\textbf {\bibinfo {volume} {15}},\ \bibinfo {pages} {2038}
  (\bibinfo {year} {2016})}\BibitemShut {NoStop}%
\bibitem [{\citenamefont {Goloborodko}\ \emph {et~al.}(2016)\citenamefont
  {Goloborodko}, \citenamefont {Imakaev}, \citenamefont {Marko},\ and\
  \citenamefont {Mirny}}]{Goloborodko2016}%
  \BibitemOpen
  \bibfield  {author} {\bibinfo {author} {\bibfnamefont {A.}~\bibnamefont
  {Goloborodko}}, \bibinfo {author} {\bibfnamefont {M.~V.}\ \bibnamefont
  {Imakaev}}, \bibinfo {author} {\bibfnamefont {J.~F.}\ \bibnamefont {Marko}},
  \ and\ \bibinfo {author} {\bibfnamefont {L.}~\bibnamefont {Mirny}},\ }\href
  {\doibase 10.7554/eLife.14864} {\bibfield  {journal} {\bibinfo  {journal}
  {eLife}\ }\textbf {\bibinfo {volume} {5}},\ \bibinfo {pages} {e14864}
  (\bibinfo {year} {2016})}\BibitemShut {NoStop}%
\bibitem [{\citenamefont {Orlandini}\ \emph {et~al.}(2019)\citenamefont
  {Orlandini}, \citenamefont {Marenduzzo},\ and\ \citenamefont
  {Michieletto}}]{Orlandini2019}%
  \BibitemOpen
  \bibfield  {author} {\bibinfo {author} {\bibfnamefont {E.}~\bibnamefont
  {Orlandini}}, \bibinfo {author} {\bibfnamefont {D.}~\bibnamefont
  {Marenduzzo}}, \ and\ \bibinfo {author} {\bibfnamefont {D.}~\bibnamefont
  {Michieletto}},\ }\href {\doibase 10.1073/pnas.1815394116} {\bibfield
  {journal} {\bibinfo  {journal} {Proceedings of the National Academy of
  Sciences}\ }\textbf {\bibinfo {volume} {116}},\ \bibinfo {pages} {8149}
  (\bibinfo {year} {2019})}\BibitemShut {NoStop}%
\bibitem [{\citenamefont {Pradhan}\ \emph {et~al.}(2024)\citenamefont
  {Pradhan}, \citenamefont {Pinto}, \citenamefont {Kanno}, \citenamefont
  {Tetiker}, \citenamefont {Baaske}, \citenamefont {Cutt}, \citenamefont
  {Chatzicharlampous}, \citenamefont {Sch{\"u}ler}, \citenamefont {Deep},
  \citenamefont {Corbett}, \citenamefont {Aragon}, \citenamefont {Virnau},
  \citenamefont {Bj{\"o}rkegren},\ and\ \citenamefont {Kim}}]{Pradhan2024}%
  \BibitemOpen
  \bibfield  {author} {\bibinfo {author} {\bibfnamefont {B.}~\bibnamefont
  {Pradhan}}, \bibinfo {author} {\bibfnamefont {A.}~\bibnamefont {Pinto}},
  \bibinfo {author} {\bibfnamefont {T.}~\bibnamefont {Kanno}}, \bibinfo
  {author} {\bibfnamefont {D.}~\bibnamefont {Tetiker}}, \bibinfo {author}
  {\bibfnamefont {M.~D.}\ \bibnamefont {Baaske}}, \bibinfo {author}
  {\bibfnamefont {E.}~\bibnamefont {Cutt}}, \bibinfo {author} {\bibfnamefont
  {C.}~\bibnamefont {Chatzicharlampous}}, \bibinfo {author} {\bibfnamefont
  {H.}~\bibnamefont {Sch{\"u}ler}}, \bibinfo {author} {\bibfnamefont
  {A.}~\bibnamefont {Deep}}, \bibinfo {author} {\bibfnamefont {K.~D.}\
  \bibnamefont {Corbett}}, \bibinfo {author} {\bibfnamefont {L.}~\bibnamefont
  {Aragon}}, \bibinfo {author} {\bibfnamefont {P.}~\bibnamefont {Virnau}},
  \bibinfo {author} {\bibfnamefont {C.}~\bibnamefont {Bj{\"o}rkegren}}, \ and\
  \bibinfo {author} {\bibfnamefont {E.}~\bibnamefont {Kim}},\ }\href {\doibase
  10.1101/2024.09.12.612694} {\bibfield  {journal} {\bibinfo  {journal}
  {bioRxiv}\ } (\bibinfo {year} {2024}),\
  10.1101/2024.09.12.612694}\BibitemShut {NoStop}%
\bibitem [{\citenamefont {Lee}\ and\ \citenamefont {Kremer}(2009)}]{Lee2009a}%
  \BibitemOpen
  \bibfield  {author} {\bibinfo {author} {\bibfnamefont {W.~B.}\ \bibnamefont
  {Lee}}\ and\ \bibinfo {author} {\bibfnamefont {K.}~\bibnamefont {Kremer}},\
  }\href {\doibase 10.1021/ma9008498} {\bibfield  {journal} {\bibinfo
  {journal} {Macromolecules}\ }\textbf {\bibinfo {volume} {42}},\ \bibinfo
  {pages} {6270} (\bibinfo {year} {2009})}\BibitemShut {NoStop}%
\bibitem [{\citenamefont {Ram{\'{i}}rez}\ \emph {et~al.}(2010)\citenamefont
  {Ram{\'{i}}rez}, \citenamefont {Sukumaran}, \citenamefont {Vorselaars},\ and\
  \citenamefont {Likhtman}}]{Ramirez2010}%
  \BibitemOpen
  \bibfield  {author} {\bibinfo {author} {\bibfnamefont {J.}~\bibnamefont
  {Ram{\'{i}}rez}}, \bibinfo {author} {\bibfnamefont {S.~K.}\ \bibnamefont
  {Sukumaran}}, \bibinfo {author} {\bibfnamefont {B.}~\bibnamefont
  {Vorselaars}}, \ and\ \bibinfo {author} {\bibfnamefont {A.~E.}\ \bibnamefont
  {Likhtman}},\ }\href {\doibase 10.1063/1.3491098} {\bibfield  {journal}
  {\bibinfo  {journal} {J . Chem. Phys.}\ }\textbf {\bibinfo {volume} {133}},\
  \bibinfo {pages} {154103} (\bibinfo {year} {2010})}\BibitemShut {NoStop}%
\bibitem [{\citenamefont {Conforto}\ \emph {et~al.}(2024)\citenamefont
  {Conforto}, \citenamefont {Gutierrez~Fosado},\ and\ \citenamefont
  {Michieletto}}]{Conforto2024}%
  \BibitemOpen
  \bibfield  {author} {\bibinfo {author} {\bibfnamefont {F.}~\bibnamefont
  {Conforto}}, \bibinfo {author} {\bibfnamefont {Y.}~\bibnamefont
  {Gutierrez~Fosado}}, \ and\ \bibinfo {author} {\bibfnamefont
  {D.}~\bibnamefont {Michieletto}},\ }\href {\doibase
  10.1103/PhysRevResearch.6.033160} {\bibfield  {journal} {\bibinfo  {journal}
  {Phys. Rev. Res.}\ }\textbf {\bibinfo {volume} {6}},\ \bibinfo {pages}
  {033160} (\bibinfo {year} {2024})}\BibitemShut {NoStop}%
\end{thebibliography}%


\begin{thebibliography}{0}%
\makeatletter
\providecommand \@ifxundefined [1]{%
 \@ifx{#1\undefined}
}%
\providecommand \@ifnum [1]{%
 \ifnum #1\expandafter \@firstoftwo
 \else \expandafter \@secondoftwo
 \fi
}%
\providecommand \@ifx [1]{%
 \ifx #1\expandafter \@firstoftwo
 \else \expandafter \@secondoftwo
 \fi
}%
\providecommand \natexlab [1]{#1}%
\providecommand \enquote  [1]{``#1''}%
\providecommand \bibnamefont  [1]{#1}%
\providecommand \bibfnamefont [1]{#1}%
\providecommand \citenamefont [1]{#1}%
\providecommand \href@noop [0]{\@secondoftwo}%
\providecommand \href [0]{\begingroup \@sanitize@url \@href}%
\providecommand \@href[1]{\@@startlink{#1}\@@href}%
\providecommand \@@href[1]{\endgroup#1\@@endlink}%
\providecommand \@sanitize@url [0]{\catcode `\\12\catcode `\$12\catcode
  `\&12\catcode `\#12\catcode `\^12\catcode `\_12\catcode `\%12\relax}%
\providecommand \@@startlink[1]{}%
\providecommand \@@endlink[0]{}%
\providecommand \url  [0]{\begingroup\@sanitize@url \@url }%
\providecommand \@url [1]{\endgroup\@href {#1}{\urlprefix }}%
\providecommand \urlprefix  [0]{URL }%
\providecommand \Eprint [0]{\href }%
\providecommand \doibase [0]{http://dx.doi.org/}%
\providecommand \selectlanguage [0]{\@gobble}%
\providecommand \bibinfo  [0]{\@secondoftwo}%
\providecommand \bibfield  [0]{\@secondoftwo}%
\providecommand \translation [1]{[#1]}%
\providecommand \BibitemOpen [0]{}%
\providecommand \bibitemStop [0]{}%
\providecommand \bibitemNoStop [0]{.\EOS\space}%
\providecommand \EOS [0]{\spacefactor3000\relax}%
\providecommand \BibitemShut  [1]{\csname bibitem#1\endcsname}%
\let\auto@bib@innerbib\@empty
\end{thebibliography}%


\begin{thebibliography}{60}

\bibitem{nasmythCohesinCatenaseSeparate2011}
Nasmyth, K. (2011) Cohesin: A catenase with separate entry and exit gates? \textit{Nat. Cell Biol.}, \textbf{13}, 1170--1177.

\bibitem{Alipour2012}
Alipour, E. and Marko, J.F. (2012) Self-organization of domain structures by DNA-loop-extruding enzymes. \textit{Nucleic Acids Res.}, \textbf{40}, 11202--11212.

\bibitem{fudenbergFormationChromosomalDomains2016}
Fudenberg, G., Imakaev, M., Lu, C., Goloborodko, A., Abdennur, N. and Mirny, L.A. (2016) Formation of Chromosomal Domains by Loop Extrusion. \textit{Cell Rep.}, \textbf{15}, 2038--2049.

\bibitem{Sanborn2015a}
Sanborn, A.L., Rao, S.S.P., Huang, S.C., Durand, N.C., Huntley, M.H., Jewett, A.I., Bochkov, I.D., Chinnappan, D., Cutkosky, A., Li, J. et al. (2015) Chromatin extrusion explains key features of loop and domain formation in wild-type and engineered genomes. \textit{Proc. Natl. Acad. Sci. USA}, \textbf{112}, E6456--E6465.

\bibitem{davidsonDNALoopExtrusion2019}
Davidson, I.F., Bauer, B., Goetz, D., Tang, W., Wutz, G. and Peters, J.M. (2019) DNA loop extrusion by human cohesin. \textit{Science}, \textbf{366}, 1338--1345.

\bibitem{ganjiRealtimeImagingDNA2018}
Ganji, M., Shaltiel, I.A., Bisht, S., Kim, E., Kalichava, A., Haering, C.H. and Dekker, C. (2018) Real-time imaging of DNA loop extrusion by condensin. \textit{Science}, \textbf{360}, 102--105.

\bibitem{Pradhan2022}
Pradhan, B., Kanno, T., Umeda Igarashi, M., Loke, M.S., Baaske, M.D., Wong, J.S.K., Jeppsson, K., Bj\"{o}rkegren, C. and Kim, E. (2023) The Smc5/6 complex is a DNA loop-extruding motor. \textit{Nature}, \textbf{616}, 843--848.

\bibitem{camaraSimpleModelExplains2021}
C{\^a}mara, A.S., Schubert, V., Mascher, M. and Houben, A. (2021) A simple model explains the cell cycle-dependent assembly of centromeric nucleosomes in holocentric species. \textit{Nucleic Acids Res.}, \textbf{49}, 9053--9065.

\bibitem{Vian2018}
Vian, L., P{\c{e}}kowska, A., Rao, S.S.P., Kieffer-Kwon, K.R., Jung, S., Baranello, L., Huang, S.C., El Khattabi, L., Dose, M., Pruett, N. et al. (2018) The Energetics and Physiological Impact of Cohesin Extrusion. \textit{Cell}, \textbf{173}, 1165--1178.

\bibitem{Gibcus2018}
Gibcus, J.H., Samejima, K., Goloborodko, A., Samejima, I., Naumova, N., Nuebler, J., Kanemaki, M.T., Xie, L., Paulson, J.R., Earnshaw, W.C. et al. (2018) A pathway for mitotic chromosome formation. \textit{Science}, \textbf{359}, eaao6135.

\bibitem{Conte2022}
Conte, M., Irani, E., Chiariello, A.M., Abraham, A., Bianco, S., Esposito, A. and Nicodemi, M. (2022) Loop-extrusion and polymer phase-separation can co-exist at the single-molecule level to shape chromatin folding. \textit{Nat. Commun.}, \textbf{13}, 4070.

\bibitem{HASSLER2018R1266}
Hassler, M., Shaltiel, I.A. and Haering, C.H. (2018) Towards a unified model of smc complex function. \textit{Curr. Biol.}, \textbf{28}, R1266--R1281.

\bibitem{Davidson2019}
Davidson, I.F., Bauer, B., Goetz, D., Tang, W., Wutz, G. and Peters, J.M. (2019) DNA loop extrusion by human cohesin. \textit{Science}, \textbf{366}, 1338--1345.

\bibitem{Ganji2018}
Ganji, M., Shaltiel, I.A., Bisht, S., Kim, E., Kalichava, A., Haering, C.H. and Dekker, C. (2018) Real-time imaging of dna loop extrusion by condensin. \textit{Science}, \textbf{360}, 102--105.

\bibitem{HIRANO2025102447}
Hirano, T. and Kinoshita, K. (2025) Smc-mediated chromosome organization: Does loop extrusion explain it all? \textit{Curr. Opin. Cell Biol.}, \textbf{92}, 102447.

\bibitem{Kim2023}
Kim, E., Barth, R. and Dekker, C. (2023) Looping the genome with smc complexes. \textit{Annu. Rev. Biochem.}, \textbf{92}, 15--41.

\bibitem{TANG20233787}
Tang, M., Pobegalov, G., Tanizawa, H., Chen, Z.A., Rappsilber, J., Molodtsov, M., Noma, K. and Uhlmann, F. (2023) Establishment of dsdna-dsdna interactions by the condensin complex. \textit{Mol. Cell}, \textbf{83}, 3787--3800.

\bibitem{Uhlmann2025}
Uhlmann, F. (2025) A unified model for cohesin function in sisterchromatid cohesion and chromatin loop formation. \textit{Mol. Cell}, \textbf{85}, 1058--1071.

\bibitem{Ryu2021}
Ryu, J.K., Bouchoux, C., Liu, H.W., Kim, E., Minamino, M., de Groot, R., Katan, A.J., Bonato, A., Marenduzzo, D., Michieletto, D. et al. (2021) Bridging-induced phase separation induced by cohesin smc protein complexes. \textit{Sci. Adv.}, \textbf{7}, eabe5905.

\bibitem{Richeldi2024}
Richeldi, M., Pobegalov, G., Higashi, T.L., Gmurczyk, K., Uhlmann, F. and Molodtsov, M.I. (2024) Mechanical disengagement of the cohesin ring. \textit{Nat. Struct. Mol. Biol.}, \textbf{31}, 23--31.

\bibitem{BONATO20215544}
Bonato, A. and Michieletto, D. (2021) Three-dimensional loop extrusion. \textit{Biophys. J.}, \textbf{120}, 5544--5552.

\bibitem{Ryu2022}
Ryu, J.K., Rah, S.H., Janissen, R., Kerssemakers, J.W.J., Bonato, A., Michieletto, D. and Dekker, C. (2022) Condensin extrudes DNA loops in steps up to hundreds of base pairs that are generated by ATP binding events. \textit{Nucleic Acids Res.}, \textbf{50}, 820--832.

\bibitem{Brackley2013pnas}
Brackley, C. A., Taylor, S., Papantonis, A.,
Cook, P. R., Marenduzzo, D. (2013) Nonspecific bridging-induced attraction drives clustering of DNA-binding proteins and genome organization. \textit{Proc. Natl. Acad. Sci. USA}, \textbf{110}, 38, e3605.


\bibitem{Chan2024pnas}
Chan, B. and Rubinstein, M. (2024) Activity-driven chromatin organization during interphase: Compaction, segregation, and entanglement suppression. \textit{Proc. Natl. Acad. Sci. USA}, \textbf{121}, e2401494121.

\bibitem{polovnikov_crumpled_2023}
Polovnikov, K.E., Brandão, H.B., Belan, S., Slavov, B., Imakaev, M. and Mirny, L.A. (2023) Crumpled Polymer with Loops Recapitulates Key Features of Chromosome Organization. \textit{Phys. Rev. X}, \textbf{13}, 041029.

\bibitem{Racko2018}
Racko, D., Benedetti, F., Goundaroulis, D. and Stasiak, A. (2018) Chromatin loop extrusion and chromatin unknotting. \textit{Polymers}, \textbf{10}, 1126.

\bibitem{Hult2017}
Hult, C., Adalsteinsson, D., Vasquez, P.A., Lawrimore, J., Bennett, M., York, A., Cook, D., Yeh, E., Forest, M.G. and Bloom, K. (2017) Enrichment of dynamic chromosomal crosslinks drive phase separation of the nucleolus. \textit{Nucleic Acids Res.}, \textbf{45}, 11159--11173.

\bibitem{Conforto2024}
Conforto, F., Gutierrez Fosado, Y. and Michieletto, D. (2024) Fluidification of entangled polymers by loop extrusion. \textit{Phys. Rev. Res.}, \textbf{6}, 033160.

\bibitem{orlandini2019synergy}
Orlandini, E., Marenduzzo, D. and Michieletto, D. (2019) Synergy of topoisomerase and structural-maintenance-of-chromosomes proteins creates a universal pathway to simplify genome topology. \textit{Proc. Natl. Acad. Sci. USA}, \textbf{116}, 8149--8154.

\bibitem{Dyson2020}
Dyson, S., Segura, J., Mart{\'{i}}nez‐Garc{\'{i}}a, B., Vald{\'{e}}s, A. and Roca, J. (2021) Condensin minimizes topoisomerase II‐mediated entanglements of DNA in vivo. \textit{EMBO J.}, \textbf{40}, e105393.

\bibitem{Nozaki2017}
Nozaki, T., Imai, R., Tanbo, M., Nagashima, R., Tamura, S., Tani, T., Joti, Y., Tomita, M., Hibino, K., Kanemaki, M.T. et al. (2017) Dynamic Organization of Chromatin Domains Revealed by Super-Resolution Live-Cell Imaging. \textit{Mol. Cell}, \textbf{67}, 282--293.

\bibitem{Gabriele2022}
Gabriele, M., Brand{\~{a}}o, H.B., Grosse-Holz, S., Jha, A., Dailey, G.M., Cattoglio, C., Hsieh, T.S., Mirny, L., Zechner, C. and Hansen, A.S. (2022) Dynamics of CTCF- and cohesin-mediated chromatin looping revealed by live-cell imaging. \textit{Science}, \textbf{376}, abn6583.

\bibitem{Hibino2024}
Hibino, K., Sakai, Y., Tamura, S., Takagi, M., Minami, K., Natsume, T., Shimazoe, M.A., Kanemaki, M.T., Imamoto, N. and Maeshima, K. (2024) Single-nucleosome imaging unveils that condensins and nucleosome–nucleosome interactions differentially constrain chromatin to organize mitotic chromosomes. \textit{Nat. Commun.}, \textbf{15}, 7152.

\bibitem{Kakui2020}
Kakui, Y., Barrington, C., Barry, D.J., Gerguri, T., Fu, X., Bates, P.A., Khatri, B.S. and Uhlmann, F. (2020) Fission yeast condensin contributes to interphase chromatin organization and prevents transcription-coupled dna damage. \textit{Genome Biol.}, \textbf{21}, 2183.

\bibitem{Bailey2023}
Bailey, M.L.P., Surovtsev, I., Williams, J.F., Yan, H., Yuan, T., Li, K., Duseau, K., Mochrie, S.G.J. and King, M.C. (2023) Loops and the activity of loop extrusion factors constrain chromatin dynamics. \textit{Mol. Biol. Cell}, \textbf{34}, ar78.

\bibitem{Mach2022}
Mach, P., Kos, P.I., Zhan, Y., Cramard, J., Gaudin, S., T\"{u}nnermann, J., Marchi, E., Eglinger, J., Zuin, J., Kryzhanovska, M. et al. (2022) Cohesin and ctcf control the dynamics of chromosome folding. \textit{Nat. Genet.}, \textbf{54}, 1907–1918.

\bibitem{Iida2025}
Iida, S., Shimazoe, M.A., Minami, K., Tamura, S., Ashwin, S.S., Higashi, K., Nishiyama, T., Kanemaki, M.T., Sasai, M., Schermelleh, L. et al. (2025) Cohesin prevents local mixing of condensed euchromatic domains in living human cells. \textit{bioRxiv}, 10.1101/2025.08.27.672592.

\bibitem{Shaltiel2022}
Shaltiel, I.A., Datta, S., Lecomte, L., Hassler, M., Kschonsak, M., Bravo, S., Stober, C., Ormanns, J., Eustermann, S. and Haering, C.H. (2022) A hold-and-feed mechanism drives directional dna loop extrusion by condensin. \textit{Science}, \textbf{376}, 1087--1094.

\bibitem{Vanderlinden2014}
Vanderlinden, W., Lipfert, J., Demeulemeester, J., Debyser, Z. and {De Feyter}, S. (2014) Structure, mechanics, and binding mode heterogeneity of LEDGF/p75-DNA nucleoprotein complexes revealed by scanning force microscopy. \textit{Nanoscale}, \textbf{6}, 4611--4619.

\bibitem{Mason2000}
Mason, T.G. (2000) Estimating the viscoelastic moduli of complex fluids using the generalized Stokes-Einstein equation. \textit{Rheol. Acta}, \textbf{39}, 371--378.

\bibitem{Kremer1990}
Kremer, K. and Grest, G.S. (1990) Dynamics of entangled linear polymer melts: A molecular-dynamics simulation. \textit{J. Chem. Phys.}, \textbf{92}, 5057--5086.

\bibitem{Plimpton1995}
Plimpton, S. (1995) Fast Parallel Algorithms for Short-Range Molecular Dynamics. \textit{J. Comp. Phys.}, \textbf{117}, 1--19.

\bibitem{Orlandini2019}
Orlandini, E., Marenduzzo, D. and Michieletto, D. (2019) Synergy of topoisomerase and structural-maintenance-of-chromosomes proteins creates a universal pathway to simplify genome topology. \textit{Proc. Natl. Acad. Sci. USA}, \textbf{116}, 8149--8154.

\bibitem{Lawrimore2016ChromoShake}
Lawrimore, J., Aicher, J.K., Hahn, P., Fulp, A., Kompa, B., Vicci, L., Falvo, M., Taylor, R.M. and Bloom, K. (2016) Chromoshake: a chromosome dynamics simulator reveals that chromatin loops stiffen centromeric chromatin. \textit{Mol. Biol. Cell}, \textbf{27}, 153--166.

\bibitem{Ramirez2010}
Ram{\'{i}}rez, J., Sukumaran, S.K., Vorselaars, B. and Likhtman, A.E. (2010) Efficient on the fly calculation of time correlation functions in computer simulations. \textit{J. Chem. Phys.}, \textbf{133}, 154103.

\bibitem{Kschonsak2017}
Kschonsak, M., Merkel, F., Bisht, S., Metz, J., Rybin, V., Hassler, M. and Haering, C.H. (2017) Structural Basis for a Safety-Belt Mechanism That Anchors Condensin to Chromosomes. \textit{Cell}, \textbf{171}, 588--600.

\bibitem{Goloborodko2016}
Goloborodko, A., Imakaev, M.V., Marko, J.F. and Mirny, L. (2016) Compaction and segregation of sister chromatids via active loop extrusion. \textit{eLife}, \textbf{5}, e14864.

\bibitem{Banigan2019}
Banigan, E.J. and Mirny, L.A. (2019) Limits of chromosome compaction by loop-extruding motors. \textit{Phys. Rev. X}, \textbf{9}, 031007.

\bibitem{Harnett2024}
Harnett, J., Weir, S. and Michieletto, D. (2024) Effects of monovalent and divalent cations on the rheology of entangled dna. \textit{Soft Matter}, \textbf{20}, 3980--3986.

\bibitem{Fosado2023Fluidification}
Fosado, Y.A.G., Howard, J., Weir, S., Noy, A., Leake, M.C. and Michieletto, D. (2023) Fluidification of entanglements by a dna bending protein. \textit{Phys. Rev. Lett.}, \textbf{130}, 058203.

\bibitem{Mason1995}
Mason, T.G. and Weitz, D.A. (1995) Optical measurements of frequency-dependent linear viscoelastic moduli of complex fluids. \textit{Phys. Rev. Lett.}, \textbf{74}, 1250--1253.

\bibitem{Lee2009a}
Lee, W.B. and Kremer, K. (2009) Entangled Polymer Melts: Relation between Plateau Modulus and Stress Autocorrelation Function. \textit{Macromolecules}, \textbf{42}, 6270--6276.

\bibitem{Martinez-Garcia2022}
Martínez-García, B., Dyson, S., Segura, J., Ayats, A., Cutts, E.E., Gutierrez-Escribano, P., Aragón, L. and Roca, J. (2022) Condensin pinches a short negatively supercoiled dna loop during each round of atp usage. \textit{EMBO J.}, \textbf{42}, e111913.

\bibitem{Rybenkov1993DNAdiameter}
Rybenkov, V.V., Cozzarelli, N.R. and Vologodskii, A.V. (1993) Probability of dna knotting and the effective diameter of the dna double helix. \textit{Proc. Natl. Acad. Sci. USA}, \textbf{90}, 5307--5311.

\bibitem{Nozaki2023}
Nozaki, T., Shinkai, S., Ide, S., Higashi, K., Tamura, S., Shimazoe, M.A., Nakagawa, M., Suzuki, Y., Okada, Y. et al. (2023) Condensed but liquid-like domain organization of active chromatin regions in living human cells. \textit{Sci. Adv.}, \textbf{9}, adf1488.

\bibitem{Tang2023}
Tang, M., Pobegalov, G., Tanizawa, H., Chen, Z.A., Rappsilber, J., Molodtsov, M., Noma, K. and Uhlmann, F. (2023) Establishment of dsdna-dsdna interactions by the condensin complex. \textit{Mol. Cell}, \textbf{83}, 3787--3800.

\bibitem{Bonato2025}
Bonato, A., Jang, J.W., Kim, D.G., Moon, K.W., Michieletto, D. and Ryu, J.K. (2025) Spontaneously directed loop extrusion in SMC complexes emerges from broken detailed balance and anisotropic DNA search. \textit{Nucleic Acids Res.}, \textbf{53}, gkaf725.

\bibitem{sapsford2025topologically}
Sapsford, E. and Michieletto, D. (2025) Topologically-crosslinked hydrogels based on $\gamma$-cyclodextrins. \textit{Commun. Chem.}, \textbf{8}, 99.

\bibitem{Murayama2018}
Murayama, Y., Samora, C.P., Kurokawa, Y., Iwasaki, H. and Uhlmann, F. (2018) Establishment of DNA-DNA Interactions by the Cohesin Ring. \textit{Cell}, \textbf{172}, 465--477.

\bibitem{hibino2024single}
Hibino, K., Sakai, Y., Tamura, S., Takagi, M., Minami, K., Natsume, T., Shimazoe, M.A., Kanemaki, M.T., Imamoto, N. and Maeshima, K. (2024) Single-nucleosome imaging unveils that condensins and nucleosome–nucleosome interactions differentially constrain chromatin to organize mitotic chromosomes. \textit{Nat. Commun.}, \textbf{15}, 7152.

\bibitem{Sun2018}
Sun, M., Biggs, R., Hornick, J. and Marko, J.F. (2018) Condensin controls mitotic chromosome stiffness and stability without forming a structurally contiguous scaffold. \textit{Chromosome Res.}, \textbf{26}, 277--295.

\end{thebibliography}
\end{document}


\title{Supplementary Information: Yeast condensin acts as a transient intermolecular crosslinker in entangled DNA}

\author{Filippo Conforto}
\affiliation{School of Physics and Astronomy, University of Edinburgh, Peter Guthrie Tait Road, Edinburgh, EH9 3FD, UK}

\author{Antonio Valdes}
\affiliation{Department of Biochemistry and Cell Biology, Julius Maximilian University of Würzburg, 97074 Würzburg, Germany}

\author{Willem Vanderlinden}
\affiliation{School of Physics and Astronomy, University of Edinburgh, Peter Guthrie Tait Road, Edinburgh, EH9 3FD, UK}


\author{Davide Michieletto}
\affiliation{School of Physics and Astronomy, University of Edinburgh, Peter Guthrie Tait Road, Edinburgh, EH9 3FD, UK}
\affiliation{MRC Human Genetics Unit, Institute of Genetics and Cancer, University of Edinburgh, Edinburgh EH4 2XU, UK}
\affiliation{International Institute for Sustainability with Knotted Chiral Meta Matter (WPI-SKCM$^2$), Hiroshima University, Higashi-Hiroshima, Hiroshima 739-8526, Japan}

\maketitle

\section{Materials and Methods} 

\subsection*{Microrheology}
For microrheology experiments, we mix 5 $\mu$l of $\lambda$DNA at 500 ng/$\mu$l, with 1 $\mu$l of 2.1 $\mu$M yeast condensin, $1$ $\mu$l of 10x condensin reaction buffer (Tris-HCl Ph 7.5 500 mM, NaCl 250 mM, MgCl2 50 mM, DTT 10 mM), 1 $\mu$l or 10 mM ATP and 1 $\mu$l of 2 $\mu$m PEGylated polystyrene beads (Polyscience). We load the sample into a 100 $\mu$m thick sample chamber comprising a microscope slide, 100 $\mu$m layer of double-sided tape and a cover slip.  We perform experiments using an Nikon Eclipse Ts2 microscope with a 60x objective and Orca Flash 4.0 CMOS camera (Hamamatsu). We record a series of movies for 2 minutes at $\sim 100$ fps on a 1024x1024 field of view, resulting in about 500 tracks per condition.

We use TrackPy and custom-written particle-tracking codes (in Python and C++) to extract the trajectories of the diffusing beads and measure the time-averaged MSDs of the diffusing particles as a function of lag time t. We note that while the tracked trajectories are in 2D, because the samples are isotropic, we average the x and y direction as if they were independent walks, such that all MSDs shown and used to determine viscosities, are determined from the average of the MSDs in the x and y directions. We compute diffusion coefficients $D$ via linear fits to the MSDs according to $MSD = (2d)Dt$ (with d = 1, because the x and y directions from 2D tracking are averaged together). From $D$, we compute the zero-shear viscosity $\nu$ using the Stokes–Einstein equation $\eta  = k_B T/(3 \pi D a)$ with $a$ the diameter of the particles. We also compute the elastic $G^\prime$ and viscous $G^{\prime \prime}$ moduli by employing the generalised Stokes-Einstein relation~\cite{Mason2000,Harnett2024}.

\subsection*{AFM sample preparation, imaging, and image processing}

As a substrate for sample deposition, we prepared poly-L-lysine-coated mica by drop-casting 20 $\mu$L poly-L-lysine (Merck; 0.01\% w/v in autoclaved milliQ water) on freshly cleaved muscovite mica (SPI Supplies) for 30 s and subsequently rinsing the surface with 20 mL of milliQ water before drying with a gentle stream of filtered N2 gas~\cite{Vanderlinden2014}. Linear DNA (500 bp, generated by PCR from pUC19 plasmid using primers 5'-AGAGCAACTCGGTCGCCGCATA (forward) and 5'-GCTTACCATCTGGCCCCAGTGC (reverse)) was mixed at final concentrations of 0.5 ng/$\mu$L DNA and 10 nM WT condensin in aqueous buffer (50 mM Tris-HCl pH = 7.5, 25 mM NaCl, 5 mM MgCl2, 1mM DTT, 1mM ATP) and incubated at room temperature for 15 s before deposition. Deposition of the sample onto poly-L-lysine coated mica was done by dropcasting. After surface adsorption for 15 s, the sample was rinsed using milliQ water (20 mL) and subsequently dried using a gentle stream of filtered N2 gas.
For atomic force microscopy imaging, we used a Nanowizard 4 XP AFM (JPK, Berlin, Germany) in tapping mode with silicon tips (FASTSCAN-A; drive frequency, 1,400 kHz; Bruker) over fields of view of 6 $\times$ 6 $\mu$m at 4,096 $\times$ 4,096 pixels and captured at line rates of 3 Hz. AFM image processing of the raw topographic data was done using  MountainSPIP software (v10, Digital Surf) included plane-fitting with a 3rd degree polynomial, and line-by-line correction with a 4th degree polynomial.

\subsection*{Electrophoretic mobility shift assay (EMSA)}
The 6-FAM labeled 50-bp dsDNA was prepared by annealing two complementary DNA oligos (Merck, \seqsplit{5'-6-FAM-GGATACGTAACAACGCTTATGCATCGCCGCCGCTACATCCCTGAGCTGAC-3'}; \seqsplit{5'-GTCAGCTCAGGGATGTAGCGGCGGCGATGCATAAGCGTTGTTACGTATCC-3'}) in annealing buffer (50 mM Tris-HCl pH 7.5, 50 mM NaCl) at a concentration of 50 $\mu$M in a temperature gradient of 0.1 C/s from 95$^\circ$C to 4$^\circ$C. The EMSA reaction was prepared with a constant DNA concentration of 10 nM, either with dsDNA or ssDNA (Merck, 5’-6-FAM-CCACTCCGAC), and the indicated concentrations of purified protein in binding buffer (50 mM Tris-HCl pH 7.5, 50 mM KCl, 125 mM NaCl, 5mM MgCl2, 5\% Glycerol, 1 mM DTT). After 10 min incubation on ice, free DNA and DNA-protein complexes were resolved by electrophoresis for 1.5 hr at 4 V/cm, on 0.75\% (w/v) TAE-agarose gels at 4$^\circ$C. 6-FAM labeled dsDNA was detected directly on a Typhoon FLA 9,500 scanner (GE Healthcare) with excitation at 473 nm with LPB (510LP) filter setting. 

\subsection*{Fluorescence Polarisation}
Fluorescence polarization (FP) experiment was performed by mixing 20 nM of the 6-FAM labeled 50 bp dsDNA or 10mer ssDNA (see Methods EMSA) with series of protein concentrations, ranging from 0.03125 $\mu$M to 32 $\mu$M, in FP buffer (25 mM Tris-HCl pH 7.5, 100 mM NaCl, 5 mM MgCl2, 1 mM DTT, 0.05\% Tween20, 0.05 mg/ml BSA). The mix was incubated for 30 min at room temperature in order to attain equilibrium. Immediately thereafter, fluorescence polarization was recorded using 485 nm and 520 nm excitation and emission filter on a Tecan SPARK Microplate reader. The change in fluorescence polarization was then plotted as mean values of three independent replicates and the dissociation constant determined.

\subsection*{Purification of Saccharomyces cerevisiae condensin holocomplex}
All five subunits, of wild-type and ATP mutants condensin complexes, were expressed from two 2$\mu$-based high copy plasmids under the control of galactose-inducible promoters transformed into \textit{S. cerevisiae}. One plasmid contained pGAL10-YCS4 pGAL1-YCG1 TRP1 and the other pGAL7-SMC4-StrepII3 pGAL1-SMC2 pGAL1-BRN1-His12-HA3 URA3 or their ATP mutant derivates. Cultures were maintained at 30$^\circ$C in –Trp–Ura medium with 2 \% (w/v) D-glucose, transferred to –Trp–Ura medium with 2 \% (w/v) raffinose for 6 h and overexpression was induced by addition of 2 \% (w/v) D-galactose for 16 h. Cell lysates were prepared in a FreezerMill (Spex) in lysis buffer (50 mM Tris-HCl pH 7.5, 200 mM NaCl, 5 \% (v/v) glycerol, 5 mM $\beta$-mercaptoethanol, 20 mM imidazole) supplemented with cOmplete EDTA-free protease inhibitor mix (cOm–EDTA, Roche). The lysate was cleared by centrifugation at 45,000 3 gmax and loaded onto Ni-Sepharose 6FF (GE Healthcare). After washing with 30–40 column volumes (cv) lysis buffer, proteins were eluted in 7–10 cv elution buffer (lysis buffer plus 300 mM imidazole). The eluate was supplemented with 0.01 \% (v/v) Tween-20, 1 mM EDTA and 0.2 mM PMSF, incubated for about 16 h with Strep-Tactin Superflow high-capacity resin (2-1208-010, IBA) and eluted with St-elution buffer (50 mM Tris-HCl pH 7.5, 200 mM NaCl, 5 \% (v/v) glycerol, 1 mM dithiothreitol (DTT) containing 10 mM desthiobiotin (D1411, Merck). The eluate was concentrated by ultracentrifugation before size-exclusion chromatography on a Superose 6 increase 10/300 column (Cytiva) pre-equilibrated in SEC-buffer (25 mM TRIS-HCl pH 7.5, 500 mM NaCl, 1 mM DTT). Peak fractions were pooled and concentrated by ultrafiltration (Vivaspin 30,000 MWCO, Sartorius).

\subsection*{Purification of Saccharomyces cerevisiae condensin hinge}

The DNA fragments encoding yeast Smc2 residues 396-792 and yeast Smc4 residues 555-951 were inserted into a pET MCN vector by standard PCR-based cloning methods. Smc2 (396-792) with an N-terminal (His)6-TEV-tag and Smc4 (555-951) without a tag were co-expressed in the Escherichia coli Rosetta (DE3) pLysS (Merck) grown at 18$^\circ$C in 2 X TY medium. Cells were lysed by sonication at 4oC in lysis buffer (50 mM TRIS-HCl pH 7.5, 200 mM NaCl, 20 mM imidazole, 5 mM b-mercaptoethanol) containing cOmplete protease inhibitor cocktail tablets without EDTA (cOm–EDTA, Roche). The lysate was cleared by centrifugation at 45,000 3 gmax and loaded onto Ni-Sepharose 6FF (GE Healthcare). After washing with 30–40 column volumes (cv) lysis buffer, proteins were eluted in 7–10 cv elution buffer (lysis buffer plus 300 mM imidazole). The eluate was dialyzed overnight in dialysis buffer (25 mM TRIS-HCl pH 7.5, 200 mM NaCl, 1 mM DTT) at 4$^\circ$C. The dialyzed eluate was diluted with low-salt buffer (25 mM TRIS-HCl pH 7.5, 100 mM NaCl, 1 mM DTT) to a final salt con- centration of 150 mM NaCl and loaded onto a 6 mL RESOURCE Q (GE Healthcare) anion exchange column pre-equilibrated with low- salt buffer. After washing with 3–5 cv low-salt buffer, proteins were eluted by increasing NaCl concentrations to 1 M in a linear gradient of 60 mL. Peak fractions were pooled and loaded onto a Superdex 200 GL 10/300 column (GE Healthcare)  equilibrated in SEC-buffer (25 mM TRIS-HCl pH 7.5, 500 mM NaCl, 1 mM DTT). Peak fractions were pooled and concentrated by ultrafiltration (Vivaspin 30,000 MWCO, Sartorius). 

\subsection*{\dm{EMSA and FP demostrate that the hinge domain can bind ssDNA}}
\begin{figure}
    \includegraphics[width=0.5\textwidth]{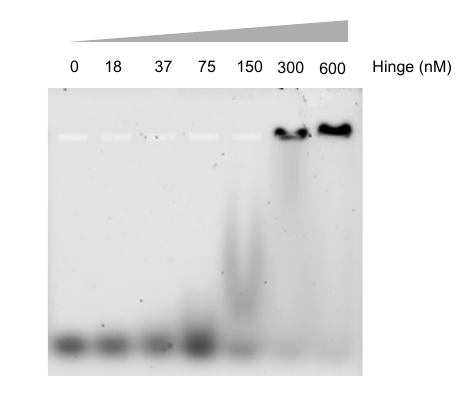}
    \caption{\dm{EMSA showing significant binding of the hinge domain
(SMC2:K841-L698, SMC4:Q646-F865) to a 10 mer ssDNA oligo in vitro. }}
    \label{fig:EMSAsi}
\end{figure}

\begin{figure}
    \includegraphics[width=0.5\textwidth]{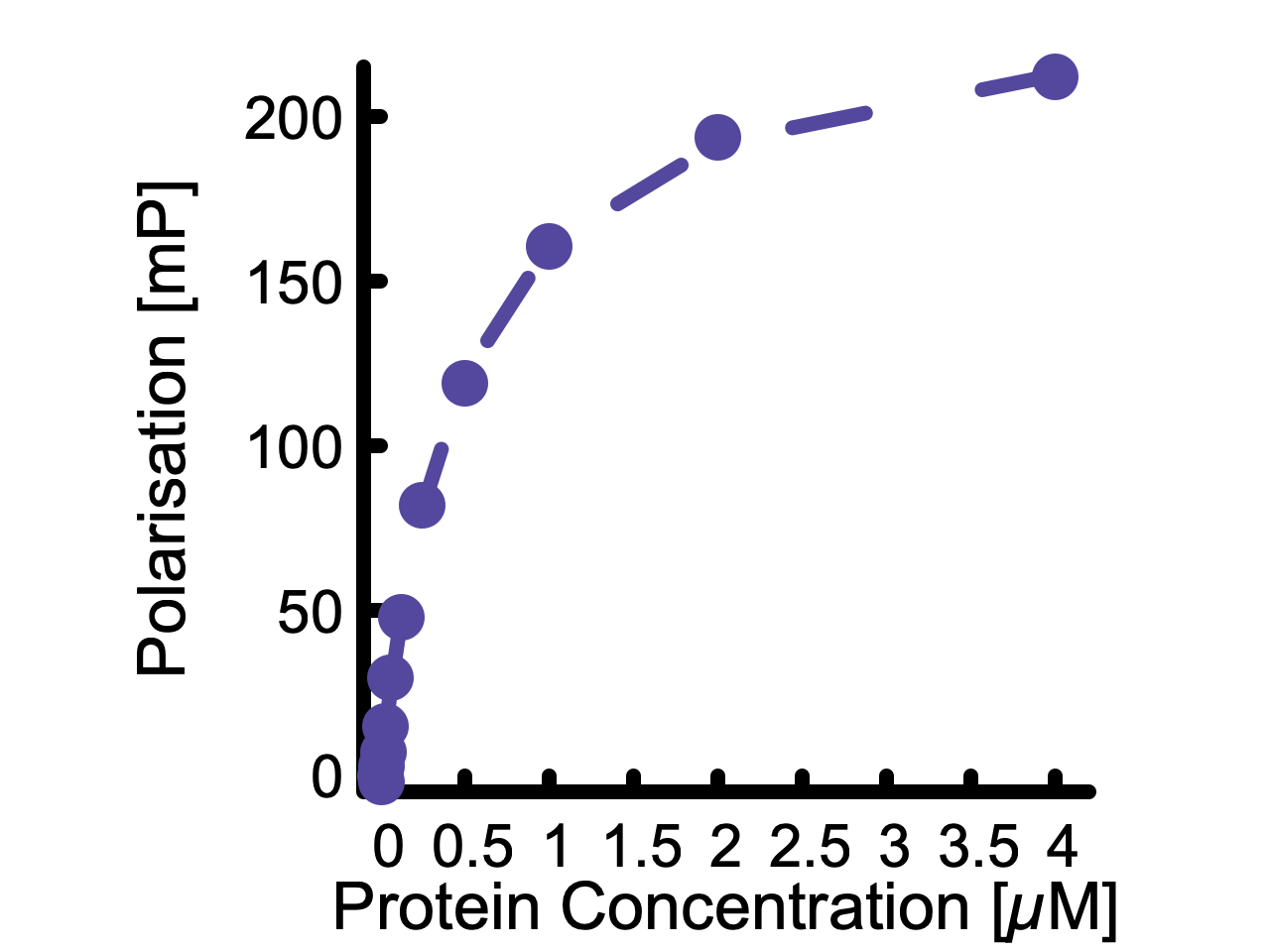}
    \caption{\dm{Fluorescence polarisation assay done with the hinge domain mixed with fluorescently-labelled ssDNA oligo of 10 mer length. Estimated $K_d = 0.48$ $\mu$M.}}
    \label{fig:FPssDNA}
\end{figure}

\dm{Through additional EMSA assays and Fluorescence Polarisation (FP) experiments we showed that the hinge 
domain of yeast condensin is able to bind both dsDNA and ssDNA. Figure~\ref{fig:EMSAsi} displays an EMSA assay done with a 10 mer ssDNA oligo, while Figure~\ref{fig:FPssDNA} shows a FP assay done with the same ssDNA oligo. Both experiments show significant binding of the hinge domain to ssDNA in vitro. The dissociation constant $k_D$ is estimated to be 0.48 from FP, similar to the value reported in the main text for dsDNA ($k_D = 0.7$).}

\subsection*{\dm{Additional microrheology experiments}}

\begin{figure}
    \centering
    \includegraphics[width=0.5\textwidth]{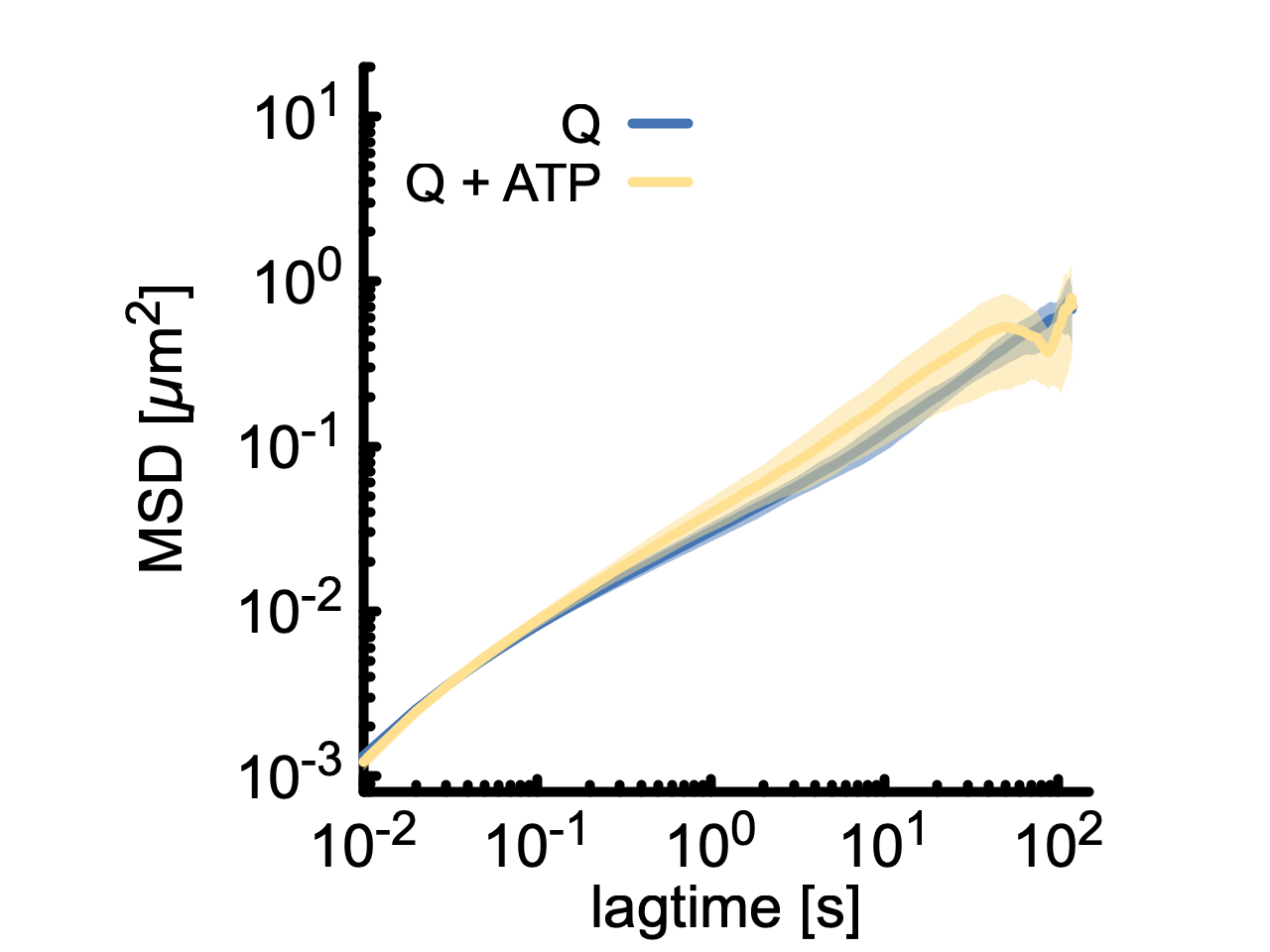}
   \includegraphics[width=0.5\textwidth]{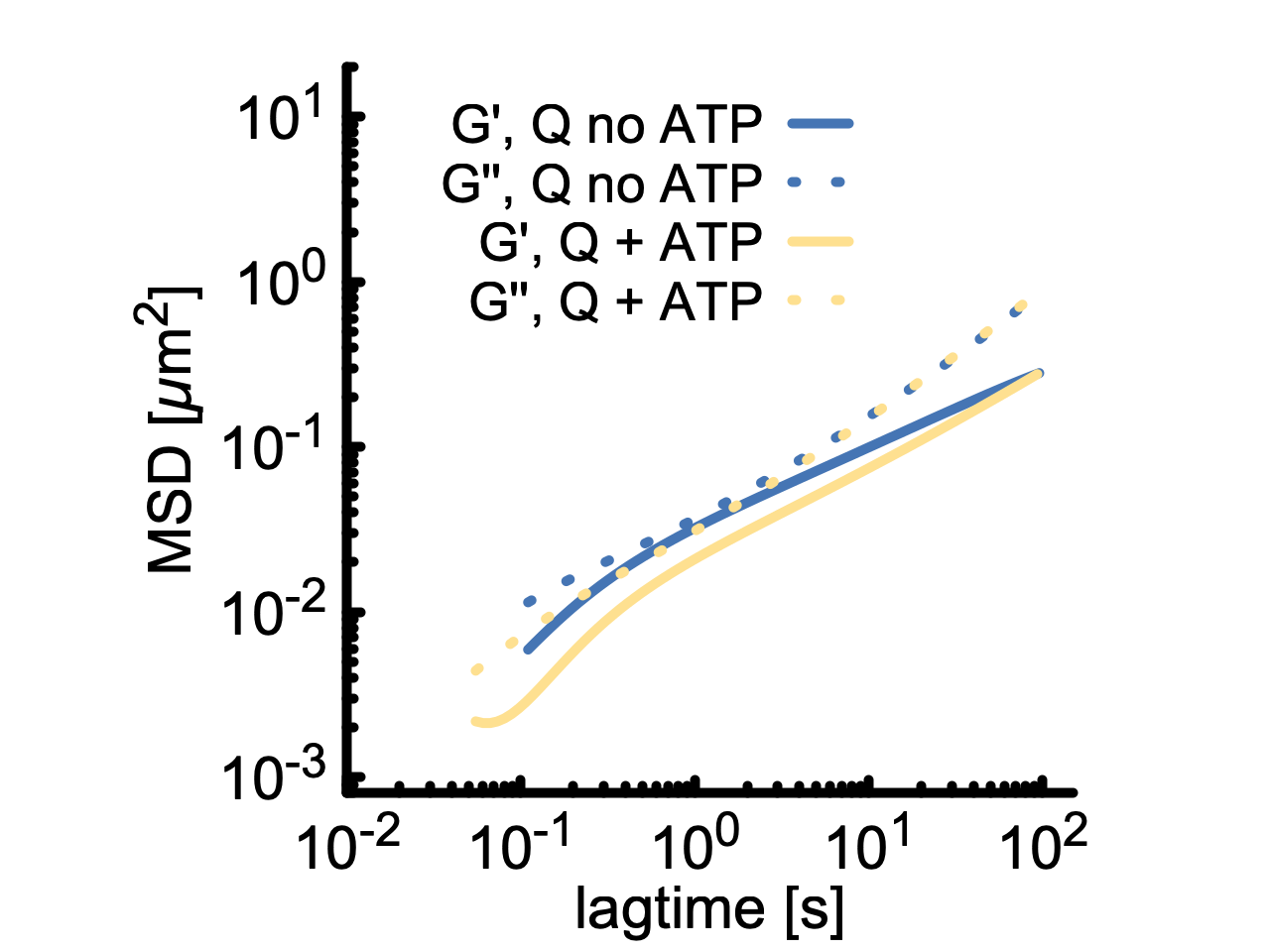}
    \caption{(Top) Mean squared displacement (MSD) of $2 \mu$m tracer beads for a solution of dense DNA (prepared as in the main text) containing Q-loop yeast mutant, in presence and absence of ATP. (bottom) G$^\prime$ and G$^{\prime \prime}$ for the samples including the Q mutant with and without ATP display similar behaviours.}
    \label{fig:fig5si}
    
\end{figure}

\dm{We performed additional microrheology experiments to assess the effect of ATP and size of passive tracer on the outcome of these measurements. 
First, we performed microrheology of a solution of $\lambda$-DNA, as in the main text, in the presence of yeast condensin Q-loop mutant (able to bind DNA but not to hydrolise ATP and therefore not able to loop extrude). We find that the MSDs of the tracers (here 2 $\mu$m particles) are similar in both presence and absence of ATP (see Fig.~\ref{fig:fig5si}). This results confirms that ATP by itself does not significantly affect the rheology of the solution or the activity of the Q-loop mutant.}

\begin{figure}[t!]
    \centering
    \includegraphics[width=0.5\textwidth]{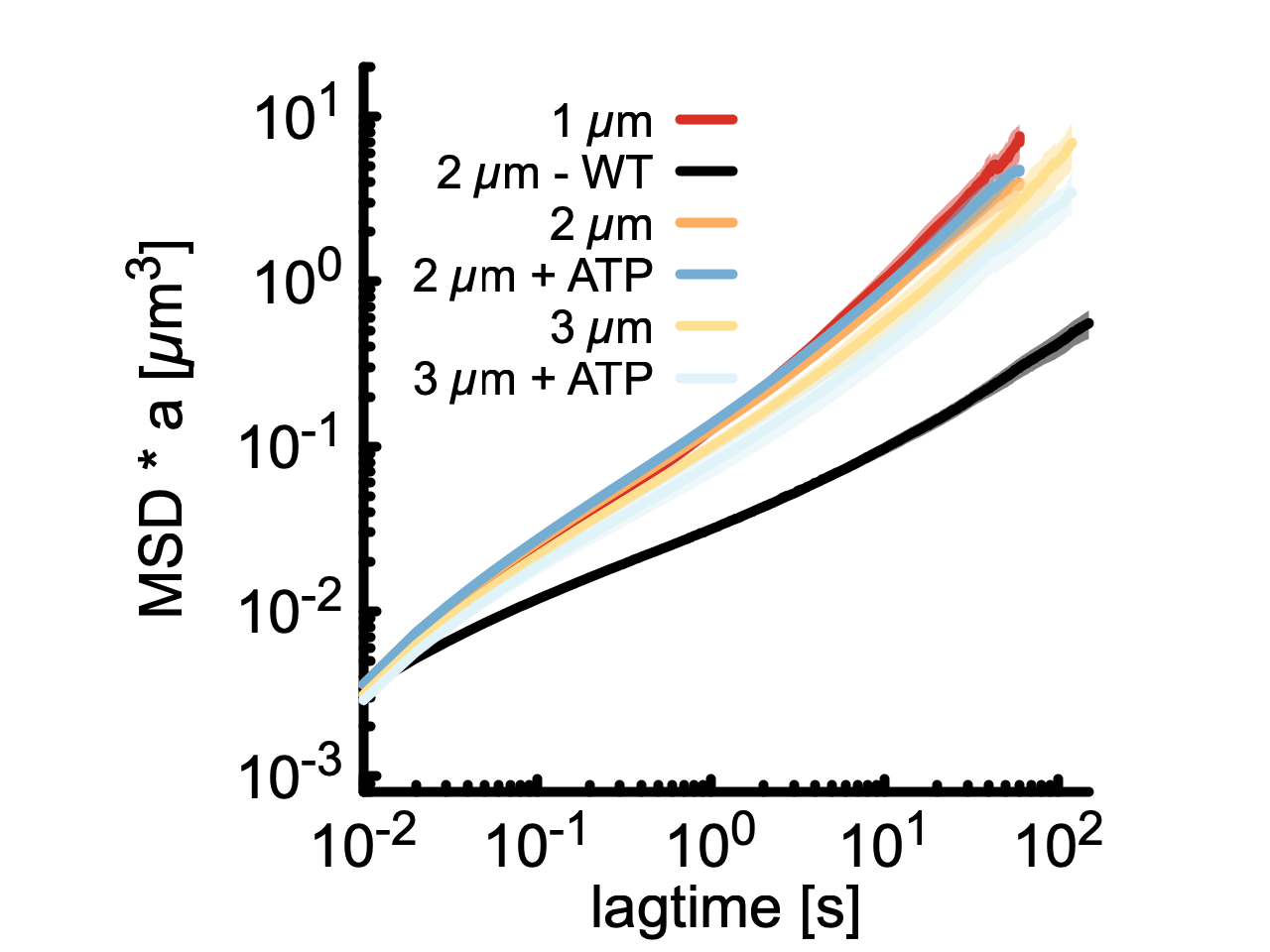}
    \caption{\dm{Mean squared displacement (MSD) multiplied by the bead size (a) computed for tracers of different sizes (1 $\mu$m, 2 $\mu$m and 3 $\mu$m) in presence or absence of ATP (no protein) and in the presence of protein and no ATP (WT).}}
    \label{fig:fig6si}
\end{figure}

\dm{Second, we performed microrheology experiments with different bead sizes to assess the role of bead size in our measurements. It is well known that while bigger tracers are better at capturing the mesoscopic rheology of the solution, small enough beads will diffuse through the entanglements and will therefore not capture mesoscale rheology~\cite{Chen2003}. Since we expect the mesh size in our experiments to be around $\simeq 0.5$ $\mu$m~\cite{Fosado2023Fluidification}, we decided to test passive tracers with diameter $> 1$ $\mu$m. We thus repeated our microrheology measurements using 1, 2 and 3 $\mu$m sized beads; we also tested different conditions: e.g. no protein and with or without ATP, to measure potential effects of Mg chelation by ATP. Because we work at low ATP concentration (1 mM) and MgCl concentration (5 mM), we do not expect significant chelation effects overall. Indeed, in fig.~\ref{fig:fig6si} we plot the MSD multiplied by the bead size (MSD $\times$ a) as a function of lag-time for experiments done with different beads and different ATP conditions. As one can appreciate, the MSD $\times$ a measured with  1 $\mu$m bead is overlapping with the one measured with the 2 $\mu$m bead in both presence and absence of ATP. The MSD $\times$ a measured with the 3 $\mu$m bead is slightly slower than expected, however the impact of bead size is not enough to significantly affect the MSD slow down observed in the presence of SMC protein (see black curve 2 $\mu$m - WT in the figure).}

\subsection*{\dm{Microrheology at different condensin concentration}}

\begin{figure}[t!]
    \centering
    \includegraphics[width=0.5\textwidth]{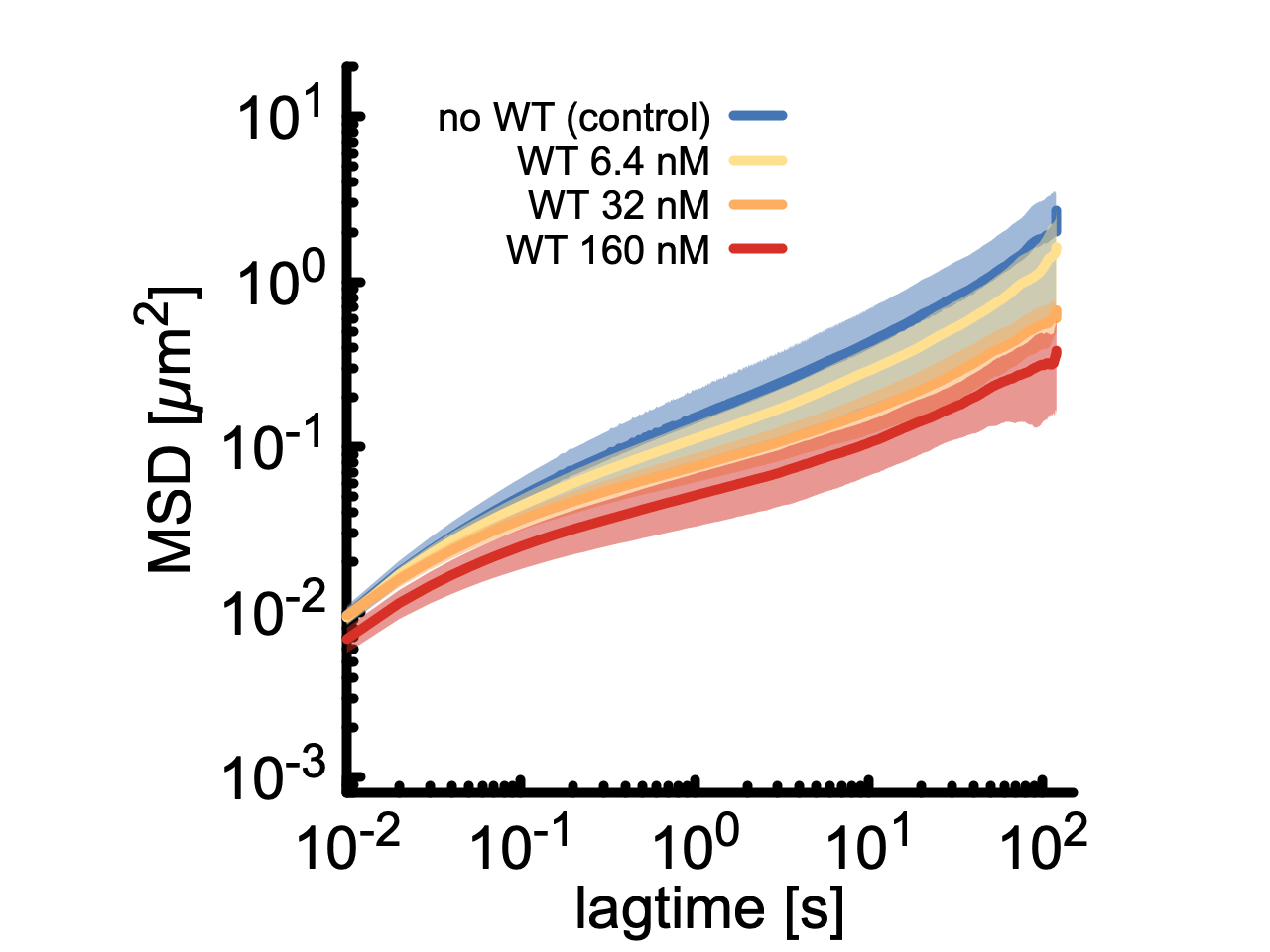}\\
      \includegraphics[width=0.5\textwidth]{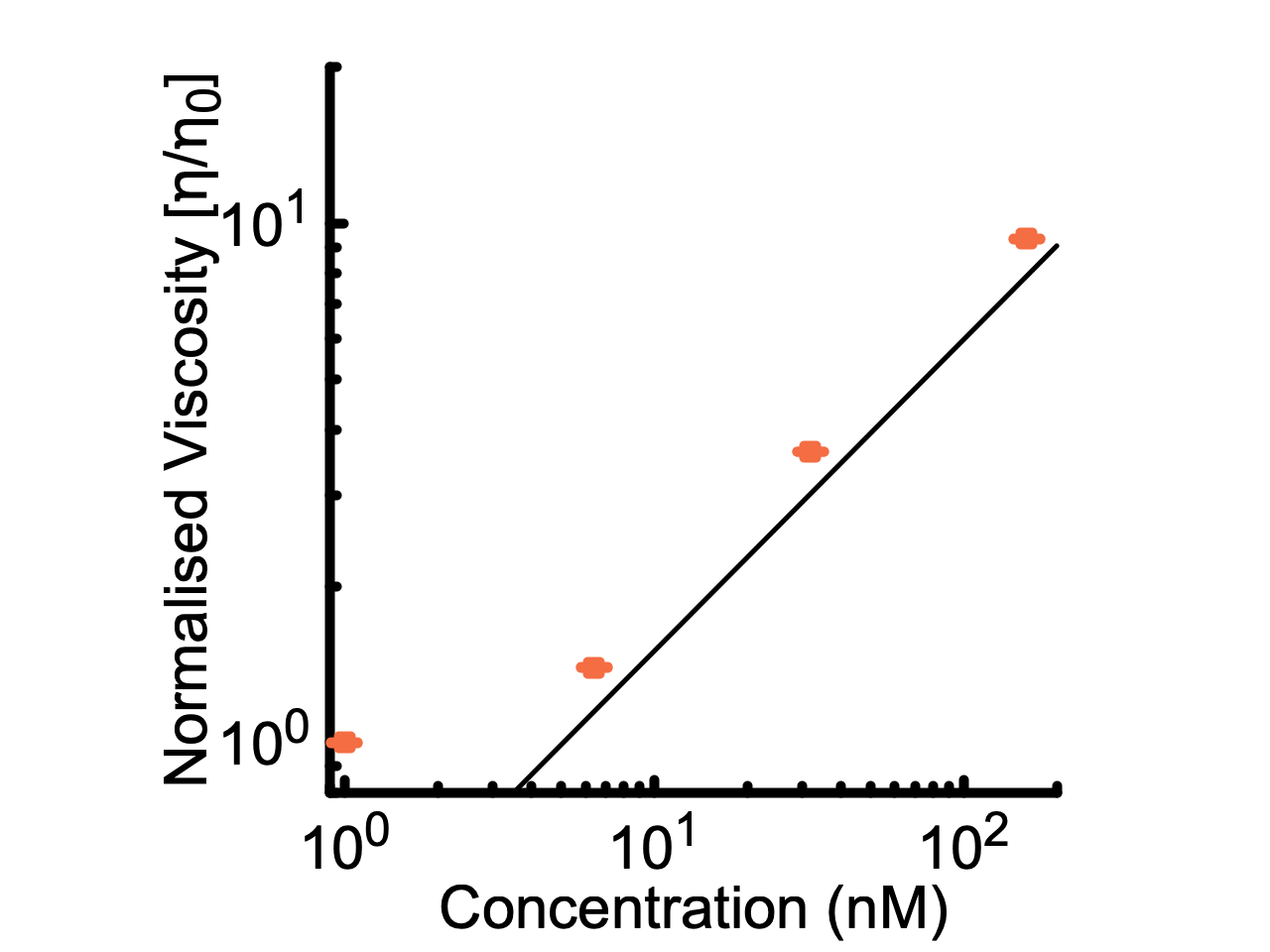}
    \caption{\dm{(Top) Mean squared displacement (MSD) of tracer beads for solutions of dense $\lambda$-DNA prepared as in the main text with different concentration of WT condensin. (Bottom) Viscosity of the solution at different SMC concentration normalised by the control (no protein) viscosity. The black line is a guide for the eye, scaling as $c^{0.6}$.}}
    \label{fig:fig3si}
\end{figure}

\dm{In this section we report microrheology experiments performed at lower condensin concentrations. The MSDs, shown in Fig.~\ref{fig:fig3si}, display an expected behaviour, at smaller protein concentration the mobility of the tracer increases, eventually matching the behaviour seen in absence of protein. The normalised viscosity scales with the concentration of SMC as $\sim c^{0.6}$. }


\subsection*{Molecular Dynamics Simulations}
We model entangled DNA as semiflexible Kremer-Grest linear polymers~\cite{Kremer1990} with $N=500$ beads of size $\sigma$. The beads interact with each other via a truncated and shifted Lennard-Jones potential,
\begin{equation}\label{eq:LJ}
    U_{\rm LJ}(r) = \left\{
    \begin{array}{lr}
        4 \epsilon \left[ \left(\dfrac{\sigma}{r}\right)^{12} - \left(\dfrac{\sigma}{r}\right)^6 + \dfrac14 \right] & \, r \le r_c \\
        0                                                                                                           & \, r > r_c
    \end{array} \right. \, ,
\end{equation}
where $r$ denotes the separation between the beads and the cut-off $r_c=2^{1/6}\sigma$ is chosen so that only the repulsive part of the potential is used. Nearest-neighbour monomers along the contour of the chains are connected by finitely extensible nonlinear elastic (FENE) springs as,
\begin{equation}
\scriptsize
\label{eq:Ufene}
        U_{\rm FENE + LJ}(r) = \left\{
        \begin{array}{lcl}
                -0.5kR_0^2 \ln\left(1-(\frac{r}{R_0})^2\right) + U_{\rm LJ} & \ r\le R_0 \\ \infty & \
                r> R_0                                                &
        \end{array} \right. \, ,
\end{equation}
where $k = 30\epsilon/\sigma^2$ is the spring constant and $R_{0}=1.5\sigma$ is the maximum extension of the elastic FENE bond. This choice of potentials and parameters is essential to preclude thermally-driven strand crossings and therefore ensures that the global topology is preserved at all times~\cite{Kremer1990,Tubiana2024}. Finally, we add bending rigidity via a Kratky-Porod potential, $U_{\rm bend}(\theta) = k_\theta \left(1 - \cos \theta \strut\right)$,  where $\theta$ is the angle formed between consecutive bonds and $k_\theta=5 k_BT$ is the bending constant, thus yielding a persistence length $l_p=5 \sigma$, corresponding to 50 nm in our grained model. 
We chose these parameters to facilitate the comparison with \textit{in vitro} experiments, i.e. to model the behaviour of naked DNA at physiological salt condition resulting in a screening length of 10 nm and a persistence length of 50 nm.  Each bead's motion is then evolved via the Langevin equation
\begin{equation}
    m \dfrac{d v_i}{d t} = - \gamma v_i  - \nabla U + \sqrt{2 k_BT \gamma_i} \eta
\end{equation}
along each Cartesian component. Here, $\gamma$ is the friction coefficient, $m$ the mass of the bead, $U$ the sum of the potentials acting on bead $i$ and $\sqrt{2k_BT \gamma} \eta$ a noise term that obeys the fluctuation-dissipation theorem, thus respecting the formula $$\langle \eta_{i}^{\alpha}(t) \eta_{j}^{\beta}(s) \rangle = \delta(t-s)\delta_{ij}\delta_{\alpha \beta}$$ along each Cartesian component (Greek letters). The numerical evolution of the Langevin equation is done with a velocity-Verlet scheme with $dt = 0.01 \tau_{LJ}$ with $\tau_{LJ} = \tau_{Br} = \sigma \sqrt{m/\epsilon}$ in LAMMPS~\cite{Plimpton1995}. 

Patches are placed at a distance of $0.5\sigma$ from the core of the beads, on the surface of each bead. Their movement follow the one of the beads they belong to. The sticky ends interaction, responsible for bridging, is modelled by a Morse potential:
$$U_m(r)=\epsilon_m \left[ e^{-2\alpha_0 (r-r_0)} - 2e^{-\alpha_0 (r-r_0)} \right]$$
for $r < R_c$. Here, $r$ represents the distance between patches of two adjacent nanostars, $r_0 = 0$ is their equilibrium distance and $R_c = 0.2\sigma$ is the cut-off distance of attraction. The amplitude of the potential is set to $\epsilon_m = 25.0 k_BT$ and $\alpha_0 = 14\sigma^{-1}$ controls the width of the potential. These parameters were chosen to ensure that during the simulation each bead can hybridize with any (but only one) of the other beads. Since the hard-sphere repulsion of beads (set by the LJ potential) covers a radius of $0.56\sigma$, we expect hybridized patches to be at a minimum distance of $0.12 \sigma$.

The system simulated consist in a solution of 50 linear polymers having size 500 in a cubic box of side length 64, achieving a monomer density of  $\sim$10\%, equivalent to a volume fraction of $\phi=0.05$. 

\subsection*{Modelling Loop Extrusion}

Our loop extrusion model was inspired by previous works \cite{fudenbergFormationChromosomalDomains2016, Goloborodko2016} which were also implemented into the LAMMPS engine. In these models, loops are formed by temporary bonds joining two beads, thus generating closed rings emerging from the polymer backbone. Loop extrusion is then achieved by shifting each bond to the adjacent beads on both sides of the bond.  However, these model allow non-physical extrusion since bond movement is performed irrespectively of the distance between selected beads. This is often possible thanks to the use of unbounded harmonic bonds, allowing large distances between the loop ends and possibly leading third segments to pass through the bonded segments. We argue that this non-physical feature should be avoided as we expect SMC complexes to block possible strand passages in between their ends, and that extrusion should take into account the geometry and topology of the DNA molecule~\cite{Orlandini2019}. Moreover, according to recent observations~\cite{Pradhan2024} extrusion has proven to be asymmetric, thus enforcing the need of pulling of only one of the two sides of the bond.
Therefore, we developed a customized version of a LAMMPS ``fix'' module publicly available at \url{https://git.ecdf.ed.ac.uk/taplab/smc-lammps} and used the version v3\_05\_07\_2024 for all the included simulations. In practice, we implement loop extrusion by initialising a given number of SMCs by choosing random triplets of beads belonging to the polymers in solution. This provides each polymer with a total number of bound SMCs on average equal to $n_{SMC}$. However, to neglect the presence of unextruded polymers we place at least one SMC on every polymer. We define the extrusion direction at the moment of the first displacement, and let the bond move and extrude a loop. Specifically, we attempt extrusion steps with a fixed frequency $f_{att}$, chosen at the beginning of the simulation. On top of that, we define a success probability $f_{prob}$ for the extrusion step, that adds up to the geometry check. Then, the distance between new SMCs' heads is computed, and the step is accepted only if its value is smaller than a fixed cut-off $r < 1.2 \sigma$.
Each SMC attempts an effective step with frequency $f_{eff} = f_{att}f_{prob}$, in turn slowing down the actual extrusion speed along the polymer because of conformational entropy. If during the extrusion process two SMCs meet on one end, extrusion stops. Consequently, extrusion runs for each SMC until the extruding end neighbours another SMC's end or reaches the polymer ends.  In the simulations we changed the value of $f_{eff} = 1 \cdot 10^{-3} \, \tau_{Br}^{-1}$ to account for different extrusion speed of the SMC complexes. Specifically we set $f_{eff} = 1 \cdot 10^{-3} \, \tau_{Br}^{-1}$, to obtain a realistic extrusion speed and evolved our simulation for the equivalent of $3 \cdot 10^6 \, \tau_{Br}$ to realise loops of different lengths. 
The latter scenario corresponds to the system defined as WT+ATP in the main text, while SMC binding and no extrusion corresponds to the case defined as WT.

To prevent integration errors related to the bond length, we initially model LEFs with a harmonic bond, with potential
\begin{equation}\label{eq:harmonic}
    U_{\rm harm}(r) = A(r-R_0) \, ,
\end{equation}
where $A=100$ and $R_0=1.1\sigma$. 
After the first step, such bond is replaced by a FENE bond with $k=10$ and $R_0=1.7\sigma$ (see Eq. 2). We choose a larger maximum extension for the elastic FENE bond and a softer spring constant to avoid bond breaking caused by sudden movement of the bonds during extrusion.
Bridging interaction is achieved by displacing uniformly (at angle of $180^\circ$ from each other if two patches, at an angle of $120^\circ$ from each other on a plane, if three patches) patchy beads on each of the beads involved in the simulation. 

\subsection{Modelling SMCs with different valence}
We can model different valence (number of interactions) by changing the number of patches on the SMC beads, as displayed in Fig.~3d in the main text. With $N_p=3$ we observed that each SMC on average makes two contacts, whilst a smaller number of patches greatly reduces the number of intermolecular contacts.

\subsection*{Green-Kubo calculation}

The stress-relaxation modulus G(t) is calculated as
\begin{equation}
G(t) = \frac{V}{3k_bT}\sum_{\alpha \neq \beta} \bar{P}_{\alpha \beta} (0) \bar{P}_{\alpha \beta} (t)\, ,\end{equation}

where ($\bar{P}_{\alpha \beta} = \bar{P}_{xy} \, \mathrm{and} \, \bar{P}_{xz} \, \mathrm{and} \, \bar{P}_{yz}$) represents the off-diagonal components of the stress tensor. Specifically, we get those components as
\begin{equation}
\bar{P}_{\alpha \beta}(t) = \frac{1}{t_{avg}}\sum_{\Delta t = -\frac{t_{avg}}{2}+1}^{t_{avg}}P_{\alpha \beta}(t+\Delta t)\, , \end{equation}
\begin{equation}
P_{\alpha \beta}(t) = \frac{1}{V}\left(\sum_{k=1}^{NM} m_k v_k^\alpha v_k^\beta +\frac{1}{2}\sum_{k=1}^{NM}\sum_{l=1}^{NM} F_{kl}^\alpha r_{kl}^\beta \right)  ,
\end{equation}
where N is the number of beads per polymer, M the number of polymers, V the box volume, $m_k$ the mass of the k-th bead, $v_k$ the speed of the k-th bead, $F_{kl}$ the force between the k-th and the l-th bead and $r_{kl}$ their distance. $P_{\alpha \beta}$ is then averaged over a time $t_{avg}$~\cite{Lee2009a}. The autocorrelation was computed using the multiple-tau correlator method described in reference~\cite{Ramirez2010} and implemented in LAMMPS with the fix ave/correlate/long command. This method makes sure that the systematic error of the multiple-tau correlator to be always below the level of the statistical error of a typical simulation (see LAMMPS documentation). The viscosity $\eta$ of the system is then obtained by integrating $G(t)$ as,
$$ \eta = \int_{0}^{t \rightarrow \infty} G(t)dt $$

Given that our simulation are run for a finite time, the computed viscosity represents a lower bound for the true value. Such value is then obtained in simulation units $\frac{k_BT\tau_{Br}}{\sigma^3}$ where $\tau_{Br}=\frac{3\pi\eta_s\sigma^3}{k_BT} = 2.3 \mu s$.

To account for the noisy values appearing on large timestep values we model the behaviour of G(t) as a stretched exponential at large times. Specifically, we define $G(t) \approx a  e^{\left(-\frac{t}{\tau}\right)^b}$, and we fit a,$\tau$,b to approximate the exponential decay, starting at an arbitrarily found point, denoted as $t_e$. The viscosity is then obtained by numerical integration up to $t_e$, while the stretched exponential contribution is obtained by computing $\int_{t_e}^{\infty} a  e^{\left(-\frac{t}{\tau}\right)^b} = \frac{a\tau}{b}\Gamma\left(\frac{1}{b},(\frac{t_e}{\tau})^b\right)$, where $\Gamma(a,z)$ is the upper generalised gamma function. The sum between these two terms returns the viscosity estimate. The fitting range is set around the length-scale at which the stress-relaxation function show the exponential decay behaviour. When this is not possible, in cases in which the relaxation timescale is longer than the duration of the simulation itself we fit the function with a power law up to $10^7 \tau_{Br}$ and use the integrated value as a lower bound estimate of the viscosity. The errors shown in Fig. 3 are defined as the range between lower bound of viscosity defined by integrating numerically up to the the fitting range and the upper bound obtained by integrating numerically up to the relaxation timescale.

The Green-Kubo (GK) measurements are done in equilibrium in all the scenarios considered in the main text. 

\subsection*{Measuring contact density}
The contact densities displayed in Fig. 3d are obtained by measuring the average number of neighbours each SMC has in the simulation over a time equal to $1 \cdot 10^6 \tau_{Br}$. We define as neighbours the beads whose patches are found to be distant from the SMC beads less than a cutoff defined as $0.2 \sigma$. By dividing it by the number of SMC in solution we get the average number of contacts per SMC, with an error that is below 1\%. Despite this consistency in the number of contacts, by fitting the exponential decay of the autocorrelation function we were able to estimate the average active time of the attractive morse between two different patches. Effectively this interaction allows the formation of temporary bridges of average duration equal to $73 \pm 2 \, \tau_{Br}$ for the system correspondent to WT, and values around $\approx 80 \tau_{Br}$ for all the tested systems. This means that the interaction modelled through this patchy system are effectively transient but can lead to semi-permanent links between the polymers.

\subsection*{The density of SMCs affects the rheology of the system}
Additionally, we tested the effect of different density of SMCs in our solution of polymers. We explored systems in which SMC were distributed such to have in average 50 SMC per polymer, which is 10 times larger than the systems considered in the main text. The viscoelasticity is then affected by the increase in the number of contacts between the polymers, especially the intermolecular ones favour the elastic behaviour of the solution, as visible in the two curves representing the WT behaviour.

\begin{figure*}[t!]
    \centering
    \includegraphics[width=\textwidth]{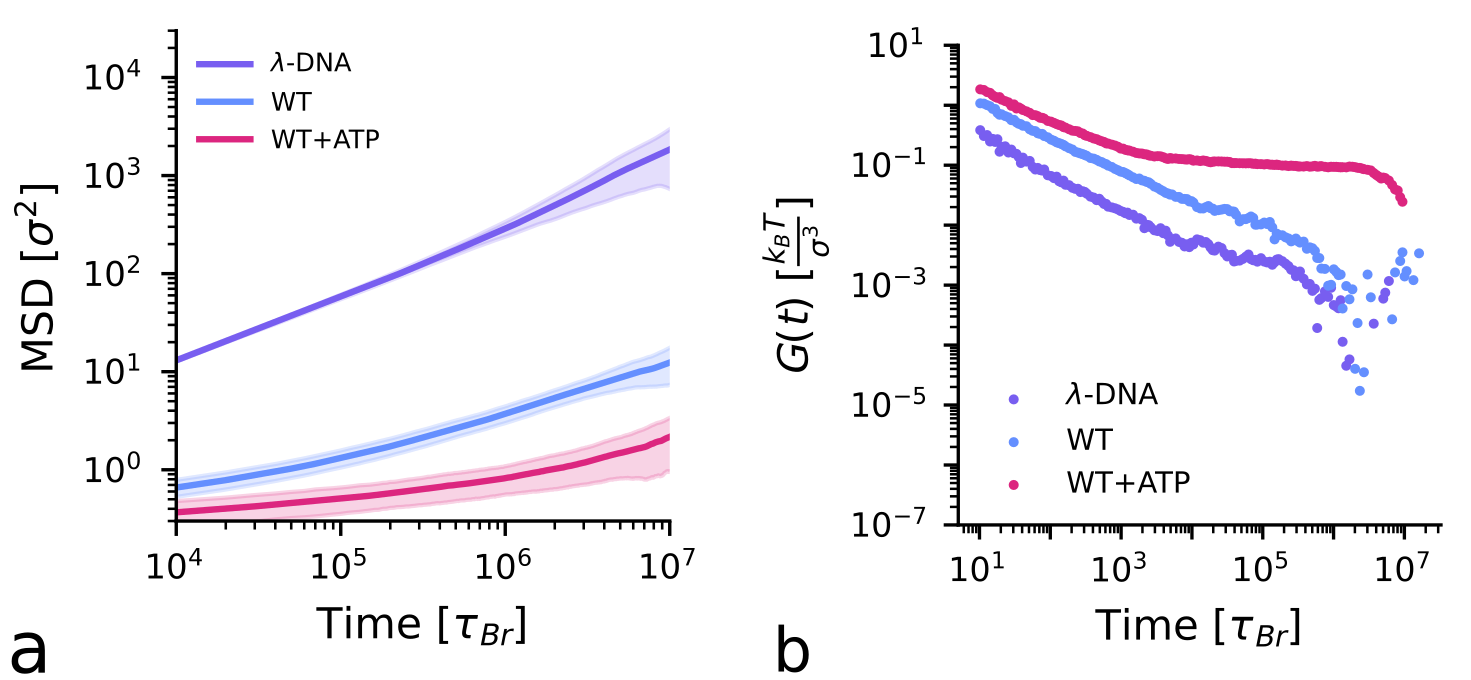}
    \caption{\textbf{High density case: 50 SMC per polymer.} \textbf{a.} Average Mean Squared Displacement (MSD) of the polymers' center of mass (standard deviation shaded) for the control case ($\lambda$-DNA) compared with the cases with SMC but no extrusion (WT) and SMC with extrusion (WT+ATP). \textbf{b.} Stress-relaxation function ($G(t)$) for the control case ($\lambda$-DNA) compared with the cases with SMC but no extrusion (WT) and SMC with extrusion (WT+ATP).}
    \label{fig:fig2si}
\end{figure*}

Surprisingly, allowing loop extrusion in such a dense system seems to favour the decrease in mobility - oppositely to what it has been reported in the main text. We argue that this effect is due to the accumulation of SMCs along the polymer. Because of the higher density and thus smaller loop size, on average shorter than the case with 5 SMC per polymer, the SMCs form clusters and thus lead to effectively stronger intermolecular bridging. As displayed in Fig.~\ref{fig:fig2si}b this clustering results in a very marked entanglement plateau in the WT+ATP case, yielding in enhanced elasticity.

\subsection{Modelling SMCs with different valence}
We can model different valence (number of interactions) by changing the number of patches on the SMC beads, as displayed in Fig.~3d in the main text. With $N_p=3$ we observed that each SMC on average makes two contacts, whilst a smaller number of patches greatly reduces the number of intermolecular contacts.

\begin{figure*}[t!]
    \centering
    \includegraphics[width=\textwidth]{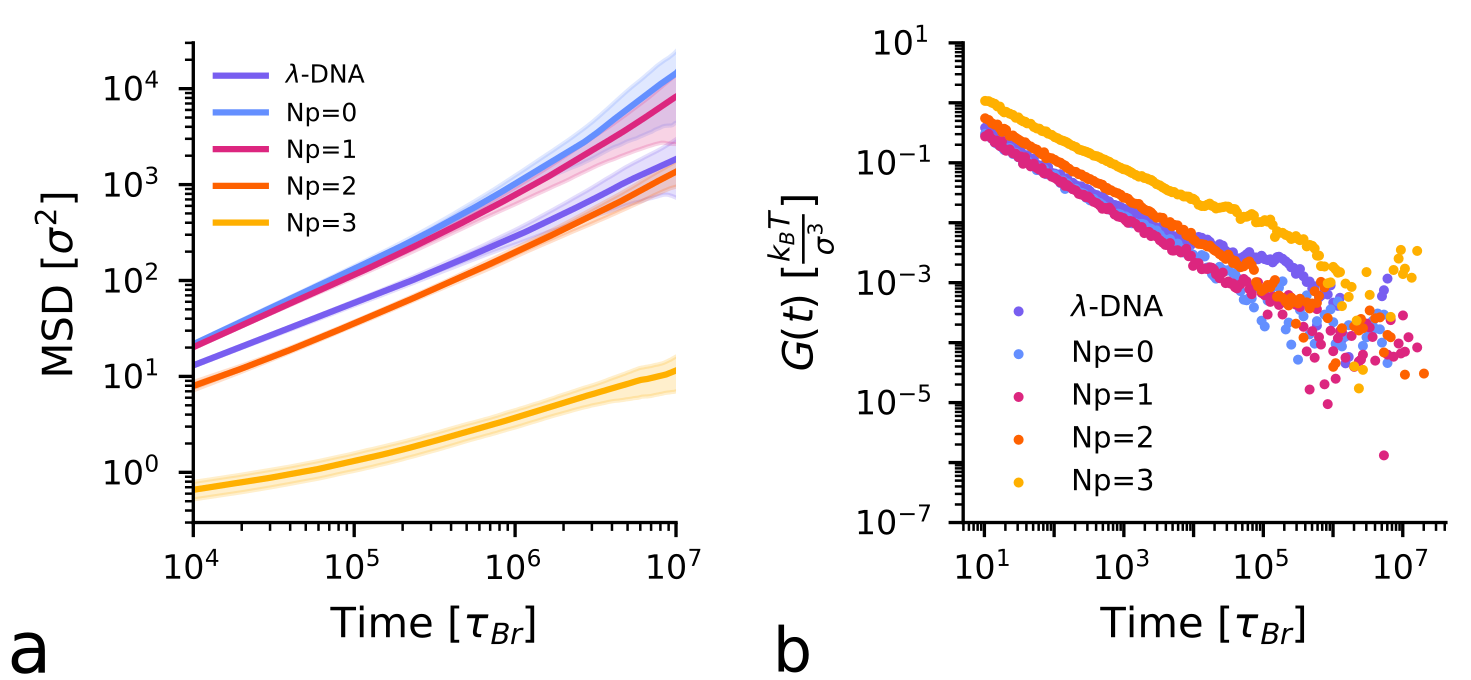}
    \caption{\textbf{Effect of number of patches on the solution dynamics.} \textbf{a.} Average Mean Squared Displacement (MSD) of the polymers' center of mass (standard deviation shaded) for the control case ($\lambda$-DNA) compared with the cases with SMC (no extrusion) and three different $N_p$. \textbf{b.} Stress-relaxation function (G(t)) of the solution of polymers for the control case ($\lambda$-DNA) compared with the cases with SMC (no extrusion) and three different $N_p$. In both these plots, the system has 50 SMCs per polymer.}
    \label{fig:fig1si}
\end{figure*}

As displayed in Fig.~\ref{fig:fig1si}a, as we increase the number of patches, the MSD of the polymers' center of mass becomes more elastic at short times. Interestingly, with $Np = 0$ and $Np = 1$ we observe an increase in mobility of the polymers, in line with what we have seen in our previous work~\cite{Conforto2024}.  Oppositely, $Np = 2$ and $Np = 3$ show significant subdiffusion; similarly, the stress relaxation function in Fig.~\ref{fig:fig1si}b shows an increase in the elastic plateau for $Np = 2$ and $Np = 3$, and the absence of an exponential decay following the elastic plateau. The integration of these curves through the fit at long times display a similar behaviour, with viscosities larger than the control for both $Np = 2$ and $Np = 3$. 

\FloatBarrier
\bibliographystyle{apsrev4-1}
\bibliography{bibliography,bib1}